\documentclass[twocolumn]{autart}
\usepackage{natbib}
\usepackage{IEEEtrantools}

\usepackage{cite}
\usepackage{booktabs} 
\usepackage{array} 
\usepackage{paralist} 
\usepackage{verbatim} 
\usepackage{mathtools}
\usepackage{mathrsfs}
\usepackage{amsfonts}
\usepackage{amsmath}
\usepackage{amssymb}
\usepackage{bbm}     

\newcommand{\ie}{i.e.,~}
\newcommand{\eg}{e.g.,~}
\renewcommand{\j}{\mathrm{j}}

\newcommand{\eiw}{\e^{\j\omega}}

\newcommand{\Tr}{\mathrm{Tr}\,}

\newcommand{\Span}{\operatorname{Span}}

\renewcommand{\Psi}{\varPsi}
\renewcommand{\Gamma}{\varGamma}
\renewcommand{\Lambda}{\varLambda}
\renewcommand{\Phi}{\varPhi}
\renewcommand{\Omega}{\varOmega}
\renewcommand{\Sigma}{\varSigma}
\renewcommand{\Theta}{\varTheta}
\renewcommand{\Pi}{\varPi}
\renewcommand{\Upsilon}{\varUpsilon}

\DeclareMathOperator{\Tvar}{tvar}
\DeclareMathOperator*{\minimize}{minimize}
\DeclareMathOperator*{\maximize}{maximize}

\newcommand{\beq}{\begin{equation}}
\newcommand{\eeq}{\end{equation}}
\newcommand{\beqn}{\begin{equation*}}
\newcommand{\eeqn}{\end{equation*}}
\newcommand{\beqarr}[1]{\begin{IEEEeqnarray}{#1}}
\newcommand{\eeqarr}{\end{IEEEeqnarray}}
\newcommand{\beqarrn}[1]{\begin{IEEEeqnarray*}{#1}}
\newcommand{\eeqarrn}{\end{IEEEeqnarray*}}
\newcommand{\bbmat}{\begin{bmatrix}}
\newcommand{\ebmat}{\end{bmatrix}}

\newcommand{\cov}{\mathrm{Cov} \: }
\newcommand{\ejw}{e^{j\omega} }

\newcommand{\diff}{\mathop{}\!\mathrm{d}}

\newcommand{\wint}[1]{\int_{-\pi}^{\pi} \!\!\!  #1 \diff \omega}
\newcommand{\expect}[1]{\mathrm{E}\left[ #1 \right]}

\newcommand{\var}[1]{\mathrm{Var} \: #1 }
\newcommand{\ascov}[1]{ \mathrm{AsCov} \: #1 }

\newcommand{\asvar}[1]{ \mathrm{AsVar} \: #1 }
\newcommand{\inp}[1]{\left \langle #1 \right\rangle}
\newcommand{\defeq}{\coloneqq} 

\DeclareMathOperator{\diag}{diag}

\newcommand{\bnul}{\begin{enumerate}[a)]}
\newcommand{\enul}{\end{enumerate}}

\newcounter{counter}
\newcommand{\upperRomannumeral}[1]{\setcounter{counter}{#1}\Roman{counter}}

\newcommand{\Bc}{{\mathcal B}}

\newcommand{\Gc}{{\mathcal G}}

\newcommand{\Lc}{{\mathcal L}}

\newcommand{\Sc}{{\mathcal S}}

\newcommand{\Cb}{{\mathbb C}}

\newcommand{\Rb}{{\mathbb R}}

\usepackage{tikz} 
\usepackage{pgfplots}
\pgfplotsset{compat=newest}
\usetikzlibrary{calc,intersections,through,backgrounds,shapes.misc,positioning,chains,scopes}
\tikzset{subsystem/.style={
rectangle,
minimum size=6mm,
thick,
draw=black,
top color=white, 
bottom color=white, 
font=\itshape,
text height=1.5ex,
text depth=0.25ex
}}

\tikzset{operator/.style={
circle,minimum size=15pt,
inner sep=0pt,
thin,draw=black, 
top color=white,bottom color=white,
text height=1.75ex,
text depth=0.25ex,
font=\itshape}}

\tikzset{signal/.style={
circle,minimum size=9pt,
inner sep=0pt,
thin,draw=black, 
top color=white,bottom color=white,
text height=1.5ex,
text depth=0.25ex,
font=\itshape}}

\tikzset{point/.style={circle,inner sep=0pt,minimum size=0pt,fill=black}}
\tikzset{skip loop/.style={to path={-- ++(0,#1) -| (\tikztotarget)}}}

\newlength\figureheight
\newlength\figurewidth
\begin{document}

\begin{frontmatter}


\title{Variance Analysis of Linear SIMO Models with Spatially Correlated Noise\thanksref{footnoteinfo}} 

\thanks[footnoteinfo]{This work was partially supported by the Swedish Research Council under contract 621-2009-4017,
and by the European Research Council under the advanced grant LEARN, contract 267381.
}

\author[kth]{Niklas Everitt}\ead{everitt@kth.se},    
\author[kth]{Giulio Bottegal}\ead{bottegal@kth.se},               
\author[kth]{Cristian R. Rojas}\ead{crro@kth.se},
\author[kth]{H{\aa}kan Hjalmarsson}\ead{hjalmars@kth.se}  

\address[kth]{ACCESS Linnaeus Center, School of Electrical Engineering, KTH - Royal Institute of Technology, Sweden}


\begin{abstract}                          
Substantial improvement in accuracy of identified linear time-invariant single-input multi-output (SIMO)
dynamical models is possible when the disturbances affecting the output measurements are spatially correlated. 
Using an orthogonal representation for the modules composing the SIMO structure, in this paper we show that the variance of a parameter estimate of a module is dependent on the model structure of the other modules, and the correlation structure of the disturbances. 
In addition, we quantify the variance-error for the parameter estimates for finite model orders, 
where the effect of noise correlation structure, model structure and signal spectra are visible. 
From these results, we derive the noise correlation structure under which the mentioned model parameterization gives the lowest variance, when one module is identified using less parameters than the other modules. 
\end{abstract}

\end{frontmatter}

\section{Introduction}
Recently, system identification in dynamic networks has gained popularity, see \eg \citet{VandenHof2013,Dankers2013indirect,Dankers2013direct,Dankers2014,Ali2011b,Materassi2011,Torres2014,Haber2014,Gunes2014,Chiuso2012}. 
In this framework, signals are modeled as nodes in a graph and edges model transfer functions.  
To estimate a transfer function in the network, a large number of methods have been proposed.
Some have been shown to give consistent estimates, provided that a certain subset of signals is included in the identification process \citep{Dankers2013indirect,Dankers2013direct}. 
In these methods, the user has the freedom to include additional signals. However, little is known on how these signals should be chosen, and how large the potential is for variance reduction. From a theoretical point of view there are only a few results regarding specific network structures \eg \citet{hagg2011identification,Ramazi20141675,wahlberg2009variance}. 
To get a better understanding of the potential of adding available information to the identification process, we ask the fundamental questions: will, and how much, an added sensor improve the accuracy of an estimate of a certain target transfer function in the network? We shall attempt to give some answers by focusing on a special case of dynamic networks, namely single-input multi-output (SIMO) systems,
and in a wider context, dynamic networks. 
%

SIMO models are interesting in themselves. They find applications in various disciplines, such as signal processing and speech enhancement \citep{benesty2005speech}, \citep{doclo2002gsvd}, communications \citep{bertaux1999parameterized}, \citep{schmidt1986multiple}, \citep{trudnowski1998simo}, biomedical sciences \citep{mccombie2005laguerre} and structural engineering \citep{ulusoy2011system}. Some of these applications are concerned with spatio-temporal models, in the sense that the measured output can be strictly related to the location at which the sensor is placed \citep{stoica1994instrumental}, \citep{viberg1997maximum}, \citep{viberg1991sensor}. 
In these cases, it is reasonable to expect that measurements collected at locations close to each other are affected by disturbances of the same nature. In other words, noise on the outputs can be correlated; understanding how this noise correlation affects the accuracy of the estimated model is a key issue in data-driven modeling of SIMO systems.

For SIMO systems, our aim in this contribution is to quantify the model error induced by stochastic disturbances and noise in prediction error identification. We consider the situation where the true system can accurately be described by the model, i.e., the true system lies within the set of models used, and thus the bias (systematic) error is zero. Then, the model error mainly consists of the variance-error, which is caused by disturbances and noise when the model is estimated using a finite number of input-output samples. In particular, we shall quantify the variance error in terms of the noise covariance matrix, input spectrum and model structure. These quantities are also crucial in answering the questions we have posed above, namely, when and how much, adding a sensor pays off in terms of accuracy of the estimated target transfer function.

There are expressions for the model error, in terms of the asymptotic (in sample size)
(co-) variance of the estimated parameters, for a variety of identification methods for multi-output systems \citep{Ljung1999},\citep{LjungCaines}.
Even though these expressions correspond to the Cram{\'e}r-Rao lower bound,
they are typically rather opaque, in that it is difficult to discern how the model accuracy is influenced by
the aforementioned quantities. 
There is a well-known expression for the variance of an estimated frequency response function that lends itself to the kind of analysis we wish to facilitate \citep{Ljung85b,yuan1984,zhu1989black}. This formula is given in its SIMO version by (see \eg \citet{zhu2001multivariable})
\begin{IEEEeqnarray*}{rCl}
\cov \! 
\hat G(\ejw) 
 \approx \frac{n}{N}\Phi_u(\omega)^{-1} \Phi_v(\omega),
\yesnumber \label{eq:asym-model-order}
\end{IEEEeqnarray*}
where $\Phi_u$ is the input spectrum and $\Phi_v$ is the spectrum of the noise affecting the outputs. Notice that the expression is valid for large number of samples $N$ and large model order $n$. For finite model order there are mainly results for SISO models \citep{Hjalmarsson2006589,Ninness&Hjalmarsson:04c,Ninness&Hjalmarsson:05a,Ninness&Hjalmarsson:05b}
and recently, multi-input-single-output (MISO) models. 
For MISO models, the concept of connectedness \citep{Gevers2006559} gives conditions on when one input can help reduce the variance of an identified transfer function.
These results  were refined in \citet{Martensson07}.
For white inputs, it was recently shown in \citet{Ramazi20141675} that an increment in the model order of one transfer function leads to an increment in the variance of another estimated transfer function only up to a point, after which the variance levels off. 
It was also quantified how correlation between the inputs may reduce the accuracy. 
The results presented here are similar in nature to those in \citet{Ramazi20141675}, while they regard another special type of multi-variable models, namely multi-input single-output (MISO) models. Note that variance expressions for the special case of SIMO cascade structures are found in \citet{wahlberg2009variance}, \citet{everitt2013geometric}.

\subsection{Contribution of this paper}
\label{subsec:Intro_example_and_contribution}
As a motivation, let us first introduce the following two output example. 
Consider the model:
\begin{IEEEeqnarray*}{rCl}
y_1(t) &=& \theta_{1,1} u(t-1) + e_1(t),   \\
y_2(t) &=& \theta_{2,2} u(t-2) + e_2(t),	
\end{IEEEeqnarray*}
where the input $u(t)$ is white noise and $e_k, k =1,2$ is measurement noise. We consider two different types of measurement noise (uncorrelated with the input). In the first case, the noise is perfectly correlated, let us for simplicity assume that $e_1(t) = e_2(t)$. 
For the second case, $e_1(t)$ and $e_2(t)$ are independent. It turns out that in the first case we can perfectly recover the parameters $\theta_{1,1}$ and $\theta_{2,2}$, while, in the second case we do not improve the accuracy of the estimate of $\theta_{1,1}$ by also using the measurement $y_2(t)$. The reason for this difference is that, in the first case, we can construct the noise free equation 
\begin{IEEEeqnarray*}{rCl}
y_1(t) - y_2(t) &=& \theta_{1,1} u(t-1) - \theta_{2,2} u(t-2)
\end{IEEEeqnarray*}
and we can perfectly recover $\theta_{1,1}$ and $\theta_{2,2}$, while in the second case neither $y_2(t)$ nor $e_2(t)$ contain information about $e_1(t)$.

Also the model structure plays an important role for the benefit of the second sensor.
To this end, we consider a third case, where again $e_1(t) = e_2(t)$. This time, the model structure is slightly different:
\begin{IEEEeqnarray*}{rCl}
y_1(t) &=& \theta_{1,1} u(t-1) + e_1(t),   \\
y_2(t) &=& \theta_{2,1} u(t-1) + e_2(t).	
\end{IEEEeqnarray*}
In this case, we can construct the noise free equation 
\begin{IEEEeqnarray*}{rCl}
y_1(t) - y_2(t) &=& (\theta_{1,1}- \theta_{2,2}) u(t-1).
\end{IEEEeqnarray*}
The fundamental difference is that now only the difference $(\theta_{1,1}- \theta_{2,1})$ can be recovered exactly, but not the parameters $\theta_{1,1}$ and $\theta_{2,1}$ themselves. They can be identified from $y_1(t)$ and $y_2(t)$ separately, as long as $y_1$ and $y_2$ are measured. A similar consideration is made in \citet{ljung2011four}, where SIMO cascade systems are considered.

This paper will generalize these observations in the following contributions:
\begin{itemize}
\item We provide  novel expressions for the variance-error of an estimated frequency response function of a SIMO linear model in orthonormal basis form, or equivalently, in fixed denominator form \citep{Ninness1999}. The expression reveals how the noise correlation structure, model orders and input variance affect the variance-error of the estimated frequency response function. 
\item For a non-white input spectrum, we show where in the frequency spectrum the benefit of the correlation structure is focused. 
\item When one module, \ie the relationship between the input and one of the outputs, is identified using less parameters, we derive the noise correlation structure under which the mentioned model parameterization gives the lowest total variance.
\end{itemize}

The paper is organized as follows: in Section \ref{sec:problem_statement} we define the SIMO model structure under study and provide an expression for the covariance matrix of the parameter estimates. Section \ref{sec:main} contains the main results, namely a novel variance expression for LTI SIMO orthonormal basis function models.
The connection with MISO models is explored in Section~\ref{sec:connection-MISO}.
In Section \ref{sec:input spectrum}, the main results are applied to a non--white input spectrum. 
In section \ref{sec:optimal_correlation_structure} we derive the correlation structure that gives the minimum total variance, when one block has less parameters than the other blocks. Numerical experiments illustrating the application of the derived results are presented in Section \ref{sec:num_exp}. A final discussion ends the paper in Section \ref{sec:conclusions}.

\section{Problem Statement} \label{sec:problem_statement}

\begin{figure}[ht]
\begin{center}
    {\includegraphics[width=6cm]{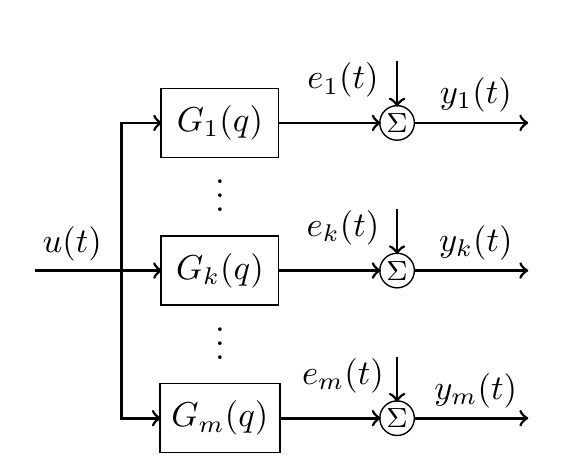}}
    \caption{\emph{Block scheme of the linear SIMO system.}} \label{fig:block_scheme}
\end{center}
\end{figure}
We consider linear time-invariant dynamic systems with one input and $m$ outputs (see Fig. \ref{fig:block_scheme}). 
The model is described as follows:
\begin{IEEEeqnarray*}{rCl}
\label{eq:model}
\begin{bmatrix}
y_1(t) \\ y_2(t) \\ \vdots \\ y_m(t)
\end{bmatrix}
&=& 
\begin{bmatrix}
G_{1}(q)  \\
G_{2}(q) 	\\
\vdots		\\
G_{m}(q) 	
\end{bmatrix}
u(t) +
\begin{bmatrix}
e_1(t) \\ e_2(t) \\ \vdots \\ e_m(t)
\end{bmatrix} \! , \yesnumber
\end{IEEEeqnarray*}
where $q$ denotes the forward shift operator, i.e., $qu(t) = u(t+1)$ and the $G_i(q)$ are causal stable rational transfer functions.  
The $G_i$ are modeled as 
\begin{IEEEeqnarray*}{rCl}
\label{eq:module_model}
G_i(q,\theta_i) =  \Gamma_i(q)\theta_i, \quad \theta_i \in \Rb^{n_i}  \! \! , \quad i = 1,\dotsc, m,
\yesnumber
\end{IEEEeqnarray*}
where $n_1 \le \dotsc \le n_m$ and
$\Gamma_i(q) =
\begin{bmatrix}
 \Bc_1(q), \dotsc, \Bc_{n_i}(q) 
 \end{bmatrix}
$, 
for some orthonormal basis functions $\{\Bc_k(q)\}_{k=1}^{n_m}$. Orthonormal with respect to 
the scalar product defined for complex functions $f(z),g(z): \Cb \to \Cb^{1 \times m}$ as
$ \inp{f,g} \defeq \frac{1}{2\pi}  \wint{  f (e^{i\omega}) g^* (e^{i\omega}) } $.
Let us introduce the vector notation 
\begin{IEEEeqnarray*}{rCl}
y(t) 
\defeq
\begin{bmatrix}
y_1(t) \\ y_2(t) \\ \vdots \\ y_m(t)
\end{bmatrix}
\! ,  \quad 
e(t) 
\defeq
\begin{bmatrix}
e_1(t) \\ e_2(t) \\ \vdots \\ e_m(t)
\end{bmatrix} \!.
\end{IEEEeqnarray*}
The noise sequence $\{e(t)\}$ is zero mean and temporally white, but may be correlated in the spatial domain:
\begin{IEEEeqnarray*}{rCl}
\label{eq:noise}
\expect{e(t)} &=& 0\\
\expect{e(t)e(s)^T} &=& \delta_{t-s} \Lambda 
\IEEEeqnarraynumspace \yesnumber
\end{IEEEeqnarray*}
for some positive definite matrix covariance matrix $\Lambda$, and where $\expect{\cdot}$ is the expected value operator. 
We express $\Lambda$ in terms of its Cholesky factorization
\begin{IEEEeqnarray*}{rCl}
\label{eq:cholesky}
\Lambda = \Lambda_{CH}\Lambda_{CH}^T,
\IEEEeqnarraynumspace \yesnumber
\end{IEEEeqnarray*}
where $\Lambda_{CH} \in \Rb^{m \times m}$ is lower triangular, i.e.,
\begin{IEEEeqnarray*}{rCl} \label{eq:gamma_def}
\Lambda_{CH} &=& 
\begin{bmatrix}
\gamma_{11} 	&	 0	 &	 \dotsc	 & 	0  \\
\gamma_{21} 	&	 \gamma_{22}	 &	 \dotsc	 & 	0  \\
\vdots 	&	 \dotsc	 &	 \ddots	 & 	0  \\
\gamma_{m1} 	&	 \gamma_{m2}	 &	 \dotsc	 & 	\gamma_{mm}  \\ 
\end{bmatrix}
\IEEEeqnarraynumspace \yesnumber
\end{IEEEeqnarray*}
for some $\{\gamma_{ij}\}$. 
Also notice that since $\Lambda > 0$, 
\begin{IEEEeqnarray*}{rCl}
\label{eq:inverse_factorization}
\Lambda^{-1} = \Lambda_{CH}^{-T} \Lambda_{CH}^{-1} .
\yesnumber
\end{IEEEeqnarray*}
We summarize the assumptions on input, noise and model as follows:
\begin{assum}
\label{ass:input_noise}
The input $\{u(t)\}$ is zero mean stationary white noise with finite moments of all orders, and variance $\sigma^2 > 0$. 
The noise $\{e(t)\}$ is zero mean and temporally white, i.e, \eqref{eq:noise} holds with $\Lambda > 0$, a positive definite matrix. It is assumed that $\expect{|e(t)|^{4 + \rho}} < \infty$ for some $\rho > 0$.  
The data is generated in open loop, that is, the input $\{u(t)\}$ is independent of the noise $\{e(t)\}$. The true input-output behavior of the data generating system can be captured by our model, i.e the true system can be described by \eqref{eq:model} and \eqref{eq:module_model} for some parameters $ \theta_i^o \in \Rb^{n_i}$, $i = 1,\dotsc,m$, where $n_1 \le \dotsc \le n_m$. 
The orthonormal basis functions $\{\Bc_k(q)\}$ are assumed stable.
$\qed$
\end{assum}

\begin{rem}
\label{rem:model_structure}
Generalization of the model structure:
\begin{itemize}
\item 
The assumption that the modules have the same orthogonal parameterization in \eqref{eq:module_model} is less restrictive than it might appear at first glance, and made for clarity and ease of presentation. A model consisting of non-orthonormal basis functions can be transformed into this format by a linear transformation, which can be computed by the Gram-Schmidt procedure \citep{trefethen1997numerical}. Notice that fixed denominator models, with the same denominator in all transfer functions, also fit this framework \citep{Ninness1999}. However, it is essential for our analysis that the modules share poles.
\item 
It is not necessary to restrict the input to be white noise. A colored input introduces a weighting by its spectrum $\Phi_u$, which means that the basis functions need to be orthogonal with respect to the inner product $\inp{f,g}_{\Phi_u} \defeq \inp{f\Phi_u,g}$. If $\Phi_u(z) =\sigma^2 R(z)R^*(z)$, where $R(z)$ is a monic stable minimum phase spectral factor; the transformation $\tilde \Gamma_i(q) =\sigma^{-1} R(q)^{-1} \Gamma_i(q)$ is a procedure that gives a set of orthogonal basis functions in the weighted space. If we use this parameterization, all the main results of the paper carry over naturally. However, in general, the new parameterization does not contain the same set of models as the original parameterization. Another way is to use the Gram-Schmidt method, which maintains the same model set and the main results are still valid.
If we would like to keep the original parameterization, we may, in some cases, proceed as in Section~\ref{sec:input spectrum} where the input signal is generated by an AR-spectrum. 
\item
Note that the assumption that $n_1 \le \dotsc \le n_m$ is not restrictive as it only represents an ordering of the modules in the system. 
\end{itemize}
\end{rem}

\subsection{Weighted least-squares estimate}

By introducing $\theta = \begin{bmatrix}
\theta_1^T, \dotsc, \theta_m^T
\end{bmatrix}^T \in \Rb^{n}
$, $n \defeq \sum_{i=1}^m n_i$ 
and the $n \times m$ transfer function matrix
\begin{IEEEeqnarray*}{rCl}
 \tilde \Psi(q) = \begin{bmatrix}
 \Gamma_1 	&	 0	 &	  	0  \\
0 &	 \ddots	 & 	0  \\
0	 &	 0	 & 	\Gamma_{m}  \\ 
\end{bmatrix} \! ,
\end{IEEEeqnarray*}
we can write the model \eqref{eq:model} as a linear regression model 
\begin{IEEEeqnarray*}{rCl}
\label{eq:regression_model}
y(t) =   \varphi^T (t)\theta + e(t),
\yesnumber
\end{IEEEeqnarray*}
where
\begin{IEEEeqnarray*}{rCl}
 \varphi^T (t) = \tilde \Psi(q)^T u(t)  .
\end{IEEEeqnarray*}
An unbiased and consistent estimate of the parameter vector $\theta$ can be obtained from weighted least-squares, with optimal weighting matrix $\Lambda^{-1}$ (see, e.g., \citet{Ljung1999,Soderstrom1989si}). $\Lambda$ is assumed known, however, this assumption is not restrictive since $\Lambda$ can be estimated from data.
The estimate of $\theta$ is given by 
\begin{IEEEeqnarray*}{rCl}
\label{eq:WLS_estimate}
\hat \theta_N &=& \left (\sum_{t=1}^N  \varphi (t) \Lambda^{-1} \varphi ^T(t) \right)^{-1}
\sum_{t=1}^N \varphi (t) \Lambda^{-1}  y(t).  
\yesnumber \IEEEeqnarraynumspace
\end{IEEEeqnarray*}
Inserting \eqref{eq:regression_model} in \eqref{eq:WLS_estimate} gives
\begin{IEEEeqnarray*}{rCl}
\hat \theta_N & = & \: \theta + \left (\sum_{t=1}^N  \varphi (t) \Lambda^{-1} \varphi^T(t) \right)^{-1}
\sum_{t=1}^N   \varphi (t) \Lambda^{-1}  e(t).
\end{IEEEeqnarray*}
Under Assumption 1, the noise sequence is zero mean, hence $\hat \theta_N $ is unbiased. 
It can be noted that this is the same estimate as the one obtained by the prediction error method and, if the noise is Gaussian, by the maximum likelihood method \citep{Ljung1999}. It also follows that the asymptotic covariance matrix of the parameter estimates is given by
\begin{IEEEeqnarray*}{rCl} \label{eq:WLS_variance}
 \ascov{\hat \theta_N} &=& \left( \expect{\varphi(t)  \Lambda^{-1} \varphi^T (t)  } \right)  ^{-1} \!\!\!\!\!\! .
 \yesnumber \IEEEeqnarraynumspace
\end{IEEEeqnarray*}
Here $\ascov \hat \theta_N$ is the asymptotic covariance matrix of the parameter estimates, in the sense that
the asymptotic covariance matrix of a stochastic sequence $\{f_N\}_{N=1}^{\infty}, f_N \in \Cb^{1 \times q}$ is defined as\footnote{This definition is slightly non-standard in that the second term is usually conjugated. For the standard definition, in general, all results have to be transposed, however, all results in this paper are symmetric.}
\begin{IEEEeqnarray*}{rCl}
\ascov f_N & \defeq & \! \lim_{N \to \infty} \! \! N \! \cdot \expect{  (f_N- \expect{f_N})^* \:\!(f_N- \expect{f_N})  } \! .
\end{IEEEeqnarray*}
In the problem we consider, using Parseval's formula and \eqref{eq:inverse_factorization}, the asymptotic covariance matrix, \eqref{eq:WLS_variance}, can be written as\footnote{Non-singularity of $\inp{\Psi,\Psi}$ usually requires parameter identifiability and persistence of excitation \citep{Ljung1999}.}
\begin{IEEEeqnarray*}{rCl}
\label{eq:inp_psi}
\ascov \hat \theta_N & = & 
\left[ \frac{1}{2\pi} \wint{\Psi(\ejw)\Psi^*(\ejw)}
\right]^{-1} 
\\
& = & \inp{\Psi,\Psi}^{-1} \! ,
\yesnumber
\end{IEEEeqnarray*}
where
\begin{IEEEeqnarray*}{rCl} \label{eq:SIMO_psi}
 \Psi(q) &=& \frac{1}{\sigma}\tilde \Psi(q)  \Lambda_{CH}^{-T}.
 \yesnumber
\end{IEEEeqnarray*} 
Note that $ \Psi(q)$ is block upper triangular since $\tilde \Psi(q)$ is block diagonal and $\Lambda_{CH}^{-T}$ is upper triangular. 


\subsection{The introductory example} 
\label{subsec:example}
With formal assumptions in place, we now consider the introductory example in greater detail.
Consider the model
\begin{IEEEeqnarray*}{rCl}
y_1(t) &=& \theta_{1,1} q^{-1}u(t) + e_1(t),   \yesnumber \label{eq:motiv_example_eq1}\\
y_2(t) &=& \theta_{2,1} q^{-1}u(t) + \theta_{2,2} q^{-2}u(t) + e_2(t)	\yesnumber \label{eq:motiv_example_eq2}
\end{IEEEeqnarray*}
%
which uses the delays $q^{-1}$ and $q^{-2}$ as orthonormal basis functions. With $\theta = [\theta_{1,1} \, \theta_{2,1} \, \theta_{2,2}]^T$; the corresponding regression matrix is
\begin{IEEEeqnarray*}{rClrCl}
\varphi(t)^T & = & \begin{bmatrix} u(t-1) & 0 & 0 \\ 0 & u(t-1) & u(t-2) \end{bmatrix} \,.
\end{IEEEeqnarray*} 
The noise vector is generated by 
\begin{IEEEeqnarray*}{rCcCl} \label{eq:noise_example}
\begin{bmatrix} e_1(t) \\ e_2(t) \end{bmatrix} 
&=& L w(t) &=&
\begin{bmatrix} 1 & 0 \\ \sqrt{1-\beta^2} & \beta  \end{bmatrix} 
\begin{bmatrix} w_1(t) \\ w_2(t) \end{bmatrix}  \,,
\yesnumber
\end{IEEEeqnarray*}
where $w_1(t)$ and $w_2(t)$ are uncorrelated white processes with unit variance. The parameter $\beta \in [0,\,1]$ tunes the correlation between $e_1(t)$ and $e_2(t)$. When $\beta = 0$, the two processes are perfectly correlated (\ie identical); conversely, when $\beta=1$, they are completely uncorrelated. Note that, for every $\beta \in [0,\,1]$, one has $\expect{e_1(t)^2} = \expect{e_2(t)^2} = 1$.
In fact, the covariance matrix of $e(t)$ becomes
\begin{IEEEeqnarray*}{rCcCl}
\label{eq:cov_e_2_by_2}
\Lambda &=& LL^T &=& \begin{bmatrix} 1 & \sqrt{1-\beta^2} \\ \sqrt{1-\beta^2} & 1 \end{bmatrix} \,.
\end{IEEEeqnarray*}
%
Then, when computing \eqref{eq:inp_psi} in this specific case gives
\begin{IEEEeqnarray*}{rCcCl}
\ascov {\hat \theta_N}   & = & \frac{1}{\sigma^2}
\begin{bmatrix} 1 & \sqrt{1-\beta^2} &  0 \\ \sqrt{1-\beta^2} & 1 & 0 \\ 0  & 0 &  \beta^2
\end{bmatrix} \,.
\yesnumber \label{eq:parameter_variance_example}
\end{IEEEeqnarray*}
We note that:
\begin{IEEEeqnarray*}{rClrCl}
\ascov{ 
\begin{bmatrix}
\hat{\theta}_{1,1} &  \hat{\theta}_{2,1}
\end{bmatrix}^T
} &=& \frac{1}{\sigma^2} \Lambda,
\\  
\ascov{ \hat{\theta}_{2,2} }
 &=& \frac{1}{\sigma^2} \beta^2.
\end{IEEEeqnarray*}

The above expressions reveals two interesting facts:
\begin{enumerate}
\item The (scalar) variances of $\hat{\theta}_{1,1}$ and $\hat{\theta}_{2,1}$, namely the estimates of parameters of the two modules related to the same time lag, are not affected by possible correlation of the noise processes, i.e., they are independent of the value of $\beta$. However, note that the cross correlation between $\hat{\theta}_{1,1}$ and $\hat{\theta}_{2,1}$ in \eqref{eq:parameter_variance_example}:
\begin{IEEEeqnarray*}{rlrCl}
    \var (\hat{\theta}_{1,1} &-  \sqrt{1-\beta^2} \hat{\theta}_{2,1})  \\
    &=\begin{bmatrix}
    1 \\ - \sqrt{1- \beta^2} 
    \end{bmatrix}^T
    \frac{1}{\sigma^2} \Lambda
    \begin{bmatrix}
    1 \\ - \sqrt{1- \beta^2}
    \end{bmatrix}
    \\
    &=    \frac{1}{\sigma^2}
    \beta^2. 
    \IEEEeqnarraynumspace \yesnumber \label{eq:parameter_cross_correlation}
\end{IEEEeqnarray*}
This cross correlation will induce a cross correlation in the transfer function estimates as well. 
\item As seen in \eqref{eq:parameter_variance_example}, the variance of $\hat{\theta}_{1,2}$ strongly depends on $\beta$. In particular, when $\beta$ tends to 0,  one is ideally able to estimate $\hat{\theta}_{1,2}$ perfectly. Note that in the limit case $\beta = 0$ one has $e_1(t) = e_2(t)$, so that \eqref{eq:model} can be rearranged to obtain the noise-free equation
\begin{IEEEeqnarray*}{rClrCl}
    y_1(t) - y_2(t)& = & (\theta_{1,1}-\theta_{2,1}) u(t-1) + \theta_{1,2}u(t-2),
\end{IEEEeqnarray*}
    which shows that both $\theta_{1,2}$ and the difference $\theta_{1,1}-\theta_{2,1}$ can be estimated perfectly. This can of course also be seen from \eqref{eq:parameter_variance_example}, cf. \eqref{eq:parameter_cross_correlation}. 
\end{enumerate}

The example shows that correlated measurements can be favorable for estimating for estimating certain parameters, but not necessarily for all. The main focus of this article is to generalize these observations to arbitrary basis functions, number of systems and number of estimated parameters. Additionally, the results are used to derive the optimal correlation structure of the noise. But first, we need some technical preliminaries.


\subsection{Review of the geometry of the asymptotic variance}
\label{sec:review_of_geom}

The following lemma is instrumental in deriving the results that follow.

\begin{lem}{(Lemma \upperRomannumeral{2}$.9$ in \citet{Hjalmar2011})}
\label{lem:ascov_as_sum_of_basis_functions}
Let $J:\Rb^{n} \to \Cb^{1 \times q}$ be differentiable with respect to $\theta$, and $\Psi \in \Lc_2^{n \times m}$; let $\Sc_{\Psi}$ be 
the subspace of $\Lc^{1 \times m}$ spanned by the rows of $\Psi$ and $\{\Bc_k^\Sc\}_{k=1}^r, r \le n$ be an orthonormal basis for $\Sc_{\Psi}$. Suppose that $J'(\theta^o) \in \Cb^{n \times q}$ is the gradient of $J$ with respect to $\theta$ and $J'(\theta^o) = \Psi(z_o)L$ for some $z_0 \in \Cb$ and $L \in \Cb^{m \times q}$. 
Then
\begin{IEEEeqnarray*}{rCl} \label{eq:ascov_as_sum_of_basis_functions}
\ascov J(\hat \theta_N)
&=&  L^*\sum_{k=1}^r \Bc_k^\Sc(z_o)^* \Bc_k^\Sc(z_o) \: L.
\yesnumber
\end{IEEEeqnarray*}
\end{lem}

\subsection{Non-estimable part of the noise}
\label{subsec:non_estimable_part}
As seen in the example of Section~\ref{subsec:example}, strong noise correlation may be helpful in the estimation. 
In fact, the variance error will depend on the non-estimable part of the noise, i.e., the part that cannot be linearly estimated from other noise sources. To be more specific, 
define the signal vector $e_{j \backslash i}(t)$ to include the noise sources from module $1$ to module $j$, with the one from module $i$ excluded, \ie
\begin{IEEEeqnarray*}{rCl}
e_{j \backslash i}(t) & \defeq & \\
&& \!\!\!\!\!\! \left\{ \begin{matrix}
\begin{bmatrix}	e_1(t), \dotsc, 	e_{j}(t)	\end{bmatrix}^T  & j < i, \\
\begin{bmatrix}	e_1(t), \dotsc, 	e_{i-1}(t)	\end{bmatrix}^T  & j = i, \\
\begin{bmatrix}	e_1(t), \dotsc, 	e_{i-1}(t), 	e_{i+1}(t), \dotsc, e_{j}(t)	\end{bmatrix}^T  & j > i. 
\end{matrix} \right. 
\end{IEEEeqnarray*}
Now, the linear minimum variance estimate of $e_i$ given $e_{j \backslash i}(t)$, is given by
\begin{IEEEeqnarray*}{rCl} \label{eq:e_est}
\hat e_{i|j}(t) \defeq \varrho_{ij}^Te_{j \backslash i}(t) .
\yesnumber \IEEEeqnarraynumspace
\end{IEEEeqnarray*}
Introduce the notation
\begin{IEEEeqnarray*}{rCl}
\lambda_{i|j}  \defeq \var [e_i(t) - \hat e_{i|j}(t) ],
\IEEEeqnarraynumspace \yesnumber \label{eq:minimum_variance_def}
\end{IEEEeqnarray*}
with the convention that $\lambda_{i|0} \defeq \lambda_i$. 
The vector $\varrho_{ij}$ in \eqref{eq:e_est} is given by  
\begin{IEEEeqnarray*}{rCl}
\varrho_{ij} = \left[ \cov{e_{j \backslash i}(t)} \right] ^{-1} \expect{e_{j \backslash i}(t) e_i(t)}.
\end{IEEEeqnarray*}
%
We call 
\begin{IEEEeqnarray*}{rCl}
e_i(t) - \hat e_{i|j}(t)
\end{IEEEeqnarray*}
the non-estimable part of $e_i(t)$ given $e_{j \backslash i}(t) $. 
\begin{defn}
When $\hat e_{i|j}(t)$ does not depend on $e_k(t)$, where $1 \le k \le j$, $k \neq i$, we say that $e_{i}(t)$ is orthogonal to $e_{k}(t)$ conditionally to $  e_{j \backslash i}(t)$.
\end{defn} 

The variance of the non-estimable part of the noise is closely related to the Cholesky factor of the covariance matrix $\Lambda$.
We have the following lemma. 
\begin{lem}
\label{lem:variance_non_estimable_part_of_noise}
Let $e(t) \in \Rb^{m}$ have zero mean and covariance matrix $\Lambda > 0$. Let $\Lambda_{CH}$ be the lower triangular Cholesky factor of $\Lambda$, i.e., $\Lambda_{CH}$ satisfies \eqref{eq:cholesky}, with $\{\gamma_{ik}\}$ as its entries as defined by \eqref{eq:gamma_def}. Then for $ j < i$,
\begin{IEEEeqnarray*}{rCl}
\lambda_{i|j}  = \sum_{k = j+1}^i \gamma_{ik}^2.
\end{IEEEeqnarray*}
Furthermore,  $\gamma_{ij} = 0$ is equivalent to that \emph{$e_{i}(t)$ is orthogonal to $e_{j}(t)$ conditionally to $  e_{j \backslash i}(t)$}.
\end{lem} 
\begin{pf}
See Appendix~\ref{sec:proof_variance_non_estimable_part_of_noise}. $\hfill \blacksquare$
\end{pf}

Similar to the above, for $i \le m$, we also define 
\begin{IEEEeqnarray*}{rCl}
e_{i:m}(t) &\defeq& 
\begin{bmatrix}	e_i(t) &  \dotsc & 	e_{m}(t)	\end{bmatrix} ^T  \! ,
\end{IEEEeqnarray*}
and for $j < i$ we define $ \hat e_{i:m|j}(t)$ as the linear minimum variance estimate of $e_{i:m}(t)$ based on the other signals $e_{j \backslash i}(t)$, and
\begin{IEEEeqnarray*}{rCl}
\Lambda_{i:m|j} &\defeq & \cov[ e_{i:m}(t) - \hat e_{i:m|j}(t) ].
\end{IEEEeqnarray*}

As a small example of why this formulation is useful, consider the covariance matrix below, where there is correlation between any pair $(e_i(t),e_j(t))$:
\begin{IEEEeqnarray*}{rCcCl}
\Lambda &=& 
\begin{bmatrix}
1 	& 0.6  & 0.9 \\
0.6 & 1    & 0.54 \\
0.9 & 0.54 & 1
\end{bmatrix} 
& = & 
\underbrace{\begin{bmatrix}
1 	& 0   & 0 \\
0.6 & 0.8 & 0 \\
0.9 & 0   & 0.44
\end{bmatrix}}_{\Lambda_{CH}}
\begin{bmatrix}
1 & 0.6  & 0.9 \\
0 & 0.8 	 & 0 \\
0 & 0 	 & 0.44
\end{bmatrix}.
\end{IEEEeqnarray*}
From the Cholesky factorization above we see that, since $\gamma_{32}$ is zero, Lemma~\ref{lem:variance_non_estimable_part_of_noise} gives that $e_3(t)$ is orthogonal to $e_2(t)$ given $e_{2 \backslash 3}(t)$, \ie there is no information about $e_3(t)$ in $e_2(t)$ if we already know $e_1(t)$. This is not apparent from $\Lambda$ where every entry is non-zero.
If we know $e_1(t)$ a considerable part of $e_2(t)$ and $e_3(t)$ can be estimated. Without knowing $e_1(t)$, $\lambda_1 = \lambda_2 = \lambda_3 = 1$, while if we know $e_1(t)$,  $\lambda_{2|1} = 0.64$ and $\lambda_{3|1} = 0.19$.

\section{Main results} \label{sec:main}

In this section, we present novel expressions for the variance-error of an estimated frequency response function. The expression reveals how the noise correlation structure, model orders and input variance affect the variance-error of the estimated frequency response function.
We will analyze the effect of the correlation structure of the noise on the transfer function estimates. 
To this end, collect all $m$ transfer functions into
\begin{IEEEeqnarray*}{rCl}
G \defeq \begin{bmatrix}
G_1	& 	G_2	&	\dotsc & G_m
\end{bmatrix}.
\end{IEEEeqnarray*}
For convenience, we will simplify notation according to the following definition:
\begin{defn} \label{def:asvar}
The asymptotic covariance of $\hat{G}(e^{j \omega_0}) \defeq G(e^{j \omega_0}, \hat \theta^N)$ for the fixed frequency $\omega_0$ is denoted by 
\begin{IEEEeqnarray*}{rCl}
 \ascov \hat G.
\end{IEEEeqnarray*} 
In particular, the variance of $\hat{G_i}(e^{j \omega_0}) \defeq G_i(e^{j \omega_0}, \hat \theta_i^N)$ for the fixed frequency $\omega_0$ is
denoted by 
\begin{IEEEeqnarray*}{rCl}
 \asvar \hat G_i.
\end{IEEEeqnarray*} 
Define $\chi_k$ as the index of the first system that contains the basis function $\Bc_{k}(e^{j\omega_0})$. Notice that $\chi_k-1$ is the number of systems that do not contain the basis function.
\end{defn} 

\begin{thm} \label{thm:cov_diag}
Let Assumption~\ref{ass:input_noise} hold. Suppose that the parameters  $\theta_i \in \Rb^{n_i}$, $i = 1, \dotsc,m$, are estimated using weighted least-squares \eqref{eq:WLS_estimate}. 
Let the entries of $\theta$ be arranged as follows:
\begin{IEEEeqnarray*}{rCl} \label{eq:theta_FIR}
        \bar \theta & =& \left[ \theta_{1,1} \, \ldots \, \theta_{m,1} \,\, \theta_{1,2} \, \ldots \, \theta_{m,2} \, \ldots \, \theta_{1,n_1} \, \ldots  \right. \nonumber \\
        &&\qquad \left. \ldots \, \theta_{m, n_1} \, \theta_{2, n_1+1} \, \ldots \, \theta_{m, n_1+1} \, \ldots \, \theta_{m,n_m} \right]^T \!.
        \IEEEeqnarraynumspace
        \yesnumber
\end{IEEEeqnarray*}
and the corresponding weighted least-squares estimate be denoted by $\hat{ \bar{\theta}}$.
Then, the covariance of $\hat{ \bar{ \theta}}$ is 
\begin{IEEEeqnarray*}{rCl} \label{eq:cor_diag}
 	\ascov \hat{ \bar{\theta}} =  \frac{1}{\sigma^2}\diag (&& \Lambda_{1:m}, \Lambda_{\chi_2:m|\chi_2-1}, \dotsc, \\ 
 	&& \: \dotsc, \Lambda_{\chi_{n_m}:m|\chi_{n_m}-1 } )  .
 	\IEEEeqnarraynumspace
 	\yesnumber
\end{IEEEeqnarray*}
In particular, the covariance of the parameters related to basis function number $k$ is given by
\begin{IEEEeqnarray*}{rCl}
 	\ascov \hat{\bar \theta}_k
 	&=&  	\frac{1}{\sigma^2} \Lambda_{\chi_k:m|\chi_k-1},
 	\IEEEeqnarraynumspace \yesnumber \label{eq:thm_ascov_theta_k}
\end{IEEEeqnarray*}
where 
\begin{IEEEeqnarray*}{rCl}
 	\hat{\bar \theta}_k &=& \begin{bmatrix}
 	\hat \theta_{\chi_k,k} \, \ldots \, \hat \theta_{m,k} 
 	\end{bmatrix}^T,
\end{IEEEeqnarray*}
 and where, for $\chi_k \le i \le m$,  
\begin{IEEEeqnarray*}{rCl}
\asvar \hat \theta_{i,k} &=& \frac{\lambda_{i|\chi_k-1}}{\sigma^2 }.
\yesnumber \label{eq:thm_var_theta_k}
\end{IEEEeqnarray*} 
It also holds that
\begin{IEEEeqnarray*}{rCl} \label{eq:thm_ascov_G}
\ascov \hat G &=& 
\sum_{k=1}^{n_m} 
\begin{bmatrix}
{\mathbf{0}}_{\chi_k-1} 	&  	\mathbf{0} \\
\mathbf{0}   &  \!\!\!\! 	\ascov \! \hat{\bar \theta}_k
\end{bmatrix}
\! |\Bc_{k}(e^{j\omega_0})|^2 \! \!
,
\IEEEeqnarraynumspace \yesnumber
\end{IEEEeqnarray*}
where $\ascov  \hat{\bar \theta}_k$ is given by \eqref{eq:thm_ascov_theta_k} and ${\mathbf{0}}_{\chi_k-1}$ is a $\chi_k-1 \times \chi_k-1$ matrix with all entries equal to zero. 
For $\chi_k=1$, ${\mathbf{0}}_{\chi_k-1}$ is an empty matrix. In \eqref{eq:thm_ascov_G}, ${\mathbf{0}}$ denotes zero matrices of dimensions compatible to the diagonal blocks.
\end{thm}

\begin{pf}
See Appendix~\ref{sec:proof_variance_G_i}. $\hfill \blacksquare$
\end{pf}

\begin{rem}
\label{rem:main_simo_remark}
The covariance of $\hat{\bar{\theta}}_{k}$, which contain the parameters related to basis function $k$, is determined by which other models share the basis function $\Bc_k$; cf. \eqref{eq:parameter_cross_correlation} of the introductory example. The asymptotic covariance of $\hat G$ can be understood as a sum of the contributions from each of the $n_m$ basis functions. The covariance contribution from a basis function $\Bc_{k}$ is weighted by $|\Bc_{k}(e^{j\omega_0})|^2$ and only affects the covariance between systems that contain that basis function, as visualized in Figure~\ref{fig:layers}.
\begin{figure}[!ht]
    \centering    
	\includegraphics[width=8cm]{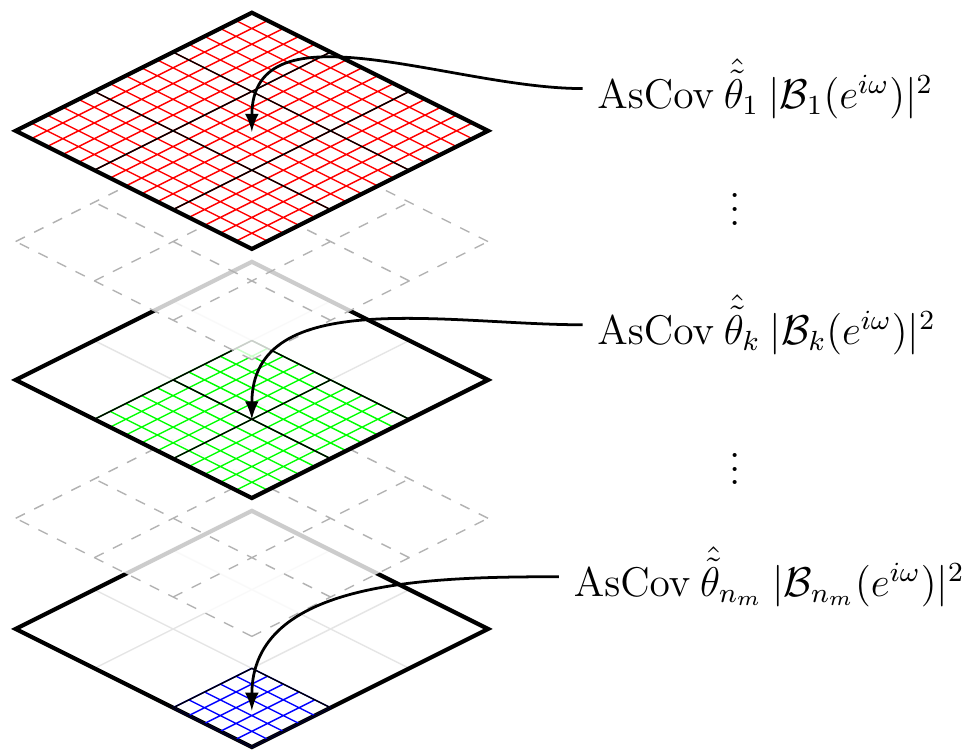}
	\caption{A graphical representation of $\ascov \hat G$ where each term of the sum in \eqref{eq:thm_ascov_G} is represented by a layer. A basis function only affects the covariance between modules that also contain that basis function. Thus, the first basis function affects the complete covariance matrix while the last basis function $n_m$ only affects modules $\chi_{n_m},\dotsc,m$.}
	 \label{fig:layers}
\end{figure} 
\end{rem}
\begin{rem}
\label{rem:theta_var_remark_explanation}
The orthogonal basis functions correspond to a decomposition of the output signals into orthogonal components and the problem in a sense becomes decoupled. As an example, consider the system described by
\begin{IEEEeqnarray*}{rCl}
 	y_1(t) &=& \theta_{1,1} \Bc_1(q)u(t) + e_1(t), \\
 	y_2(t) &=& \theta_{2,1} \Bc_1(q)u(t) + e_2(t), \\
 	y_3(t) &=& \theta_{3,1} \Bc_1(q)u(t) +  \theta_{3,2} \Bc_2(q)u(t)+ e_3(t), 
 	\IEEEeqnarraynumspace \yesnumber \label{eq:SIMO_blocks_example_1}
\end{IEEEeqnarray*}
Suppose that we are interested in estimating $\theta_{3,2}$. For this
parameter, \eqref{eq:thm_var_theta_k} becomes
\begin{IEEEeqnarray*}{rCl}
\asvar \hat{\theta}_{3,2} &=& \frac{\lambda_{3|2}}{\sigma^2} 
 	\IEEEeqnarraynumspace \yesnumber  \label{eq:SIMO:ee}
\end{IEEEeqnarray*}
To understand the mechanisms behind this expression, let $u_1(t) = {\mathcal
B}_1(q)u(t)$, and $u_2(t) = {\mathcal B}_2(q)u(t)$ so that the
system can be visualized as in Figure~\ref{fig:SIMO_blocks_example_1}, \ie we can consider $u_1$
and $u_2$ as separate inputs.
 
First we observe that it is only $y_3$ that contains information about
$\theta_{3,2}$, and the term $\theta_{3,1}u_1$ contributing to $y_3$
is a nuisance from the perspective of estimating
$\theta_{3,2}$. This term vanishes when $u_1=0$ and we will not be
able to achieve better accuracy than the optimal estimate of
$\theta_{3,2}$ for this idealized case. So let us study this setting first.
Straightforward application of
the least-squares method, using $u_2$ and $y_3$, gives an
estimate of $\theta_{3,2}$ with variance $\lambda/\sigma^2$, which is
larger than
\eqref{eq:SIMO:ee} when $e_3$ depends on $e_1$ and $e_2$. However, in this
idealized case, $y_1=e_1$ and $y_2=e_2$, and these signals can thus be
used to estimate $e_3$. This estimate can then be subtracted from
$y_3$ before the least-squares method is applied. The remaining noise
in $y_3$ will have variance $\lambda_{3|2}$, if $e_3$
is optimally  estimated (see \eqref{eq:e_est}--\eqref{eq:minimum_variance_def}), and hence the
least-squares estimate will now have variance
$\lambda_{3|2}/\sigma^2$, \ie the same as \eqref{eq:SIMO:ee}.

To understand why it is possible to achieve the same accuracy as this
idealized case when $u_1$ is non-zero, we need to observe that our new
inputs $u_1(t)$ and  $u_2(t)$ are orthogonal (uncorrelated)\footnote{This since $u(t)$ is white and
${\mathcal B}_1$ and ${\mathcal B}_2$ are orthonormal.}. Returning to
the case when only the output $y_3$ is used for estimating
$\theta_{3,2}$, this implies that we pay no price for including the
term $\theta_{3,1}u_1$ in our model, and then estimating
$\theta_{3,1}$ and $\theta_{3,2}$ jointly, \ie the variance of
$\hat{\theta}_{3,2}$ will still be $\lambda/\sigma^2$\footnote{With $u_1$ and
$u_2$ correlated, the variance will be higher, see Section~\ref{sec:connection-MISO} for a
further discussion of this topic.}. 
The question now is if we can use
$y_1$ and $y_2$ as before to estimate $e_3$. Perhaps surprisingly, we can use
the same estimate as when $u_1$ was zero. The reader may object that
this estimate will now, in addition to the previous optimal estimate of $e_3$,
contain a term which is a multiple of $u_1$. However, due to the
orthogonality between $u_1$ and $u_2$, this term will
only affect the estimate of $\theta_{3,1}$ (which we anyway were not
interested in, in this example), and the accuracy of the
estimate of $\theta_{3,2}$ will be $\lambda_{3|2}/\sigma^2$, i.e.
\eqref{eq:SIMO:ee}. Figure~\ref{fig:SIMO_blocks_example_2} illustrates the setting with $\tilde{y}_3$
denoting $y_3$ subtracted by the optimal estimate of $e_3$.
In the figure, the new parameter $\tilde{\theta}_{3,1}$ reflects that the relation between
$u_1$ and $\tilde{y}_3$ is different from $\theta_{3,1}$ as discussed above.
A key insight from this discussion is that for the estimate of a
parameter in the path from input $i$ to output $j$, it is only outputs
that are not affected by input $i$ that can be used to estimate the
noise in output $j$; when this particular parameter is estimated,
using outputs influenced by input $i$ will introduce a bias, since the
noise estimate will then contain a term that is not orthogonal to this
input. In \eqref{eq:thm_var_theta_k}, this manifests itself in that the numerator is $\lambda_{i|\chi_k-1}$, only the $\chi_k-1$ first systems do not contain $u_i$.
\begin{figure}[!ht]
    \centering    
	\includegraphics[width=6cm]{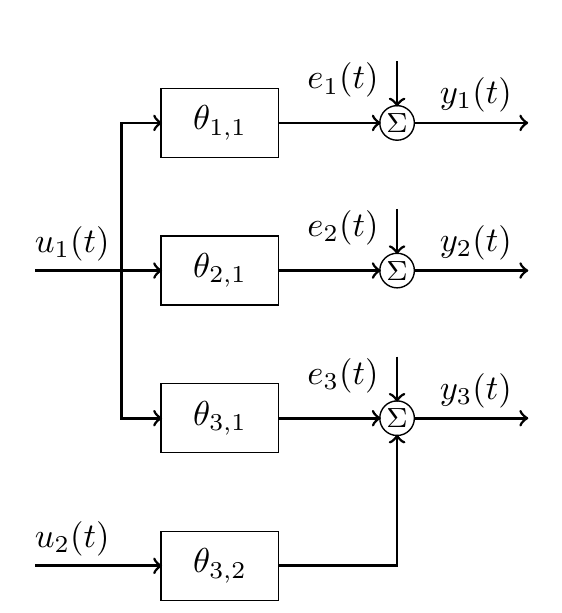}
	\caption{The SIMO system of Remark~\ref{rem:theta_var_remark_explanation}, described by \eqref{eq:SIMO_blocks_example_1}.}
	 \label{fig:SIMO_blocks_example_1}
\end{figure} 
\begin{figure}[!ht]
    \centering    
	\includegraphics[width=6cm]{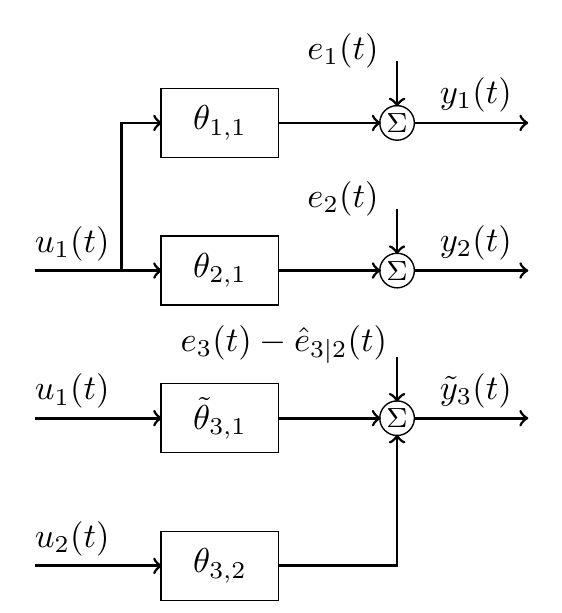}
	\caption{The SIMO system of Remark~\ref{rem:theta_var_remark_explanation}, described by \eqref{eq:SIMO_blocks_example_1}, but with  $\tilde{y}_3$ denoting $y_3$ subtracted by the optimal estimate of $e_3$ and $\tilde{\theta}_{3,1}$ reflects that the relation between
$u_1$ and $\tilde{y}_3$ is different from $\theta_{3,1}$.}
	 \label{fig:SIMO_blocks_example_2}
\end{figure}

\end{rem}

We now turn our attention to the variance of the individual transfer function estimates. 

\begin{cor}
\label{cor:variance_G_i}
Let the same assumptions as in Theorem~\ref{thm:cov_diag} hold. Then, for any frequency $\omega_0$, it holds that 
\begin{IEEEeqnarray*}{rCl}
\label{eq:ascov_G_i_hat}
\asvar \hat G_i  &=& \sum_{k=1}^{n_i} |\Bc_k(e^{j\omega_0})|^2  \asvar \hat \theta_{i,k} ,
\yesnumber
\end{IEEEeqnarray*} 
where 
\begin{IEEEeqnarray*}{rCl}
\asvar \hat \theta_{i,k} &=& \frac{\lambda_{i|\chi_k-1}}{\sigma^2 },
\yesnumber \label{eq:cor_var_theta_k}
\end{IEEEeqnarray*} 
and $\lambda_{i|j}$ is defined in \eqref{eq:minimum_variance_def}.
\end{cor}

\begin{pf}
Follows from Theorem~\ref{thm:cov_diag}, 
since \eqref{eq:ascov_G_i_hat} is a diagonal element of \eqref{eq:thm_ascov_G}. $\hfill \blacksquare$
\end{pf}

From Corollary~\ref{cor:variance_G_i}, we can tell when increasing the model order of $G_j$ will increase the asymptotic variance of $\hat G_i$. 

\begin{cor}
\label{cor:variance_increase}
Under the same conditions as in Theorem~\ref{thm:cov_diag}, 
if we increase the number of estimated parameters of $G_j$ from $n_j$ to $n_j+1$,
the asymptotic variance of $G_i$ will increase, if and only if all the following conditions hold:
\begin{enumerate}
\item $n_j< n_i$,
\item $e_i(t)$ is not orthogonal to $e_j(t)$ conditioned on $e_{j \backslash i}(t)$,
\item $ |\Bc_{n_{j+1}}(e^{j\omega_0})|^2 \neq 0$.
\end{enumerate}
\end{cor}

\begin{pf}
See Appendix~\ref{sec:proof_variance_increase}. $\hfill \blacksquare$
\end{pf}

\begin{rem}
Corollary~\ref{cor:variance_increase} explicitly tells when an increase in the model order of $G_j$ from $n_j$ to $n_j+1$ will increase the variance of $G_i$. If $n_j \ge n_i$, there will be no increase in the variance of $G_i$, no matter how many additional parameters we introduce to the model $G_j$, which was also seen the introductory example in Section~\ref{subsec:example}. Naturally, if $ e_{i}(t)$ is orthogonal to $e_j(t)$ conditioned on $e_{j \backslash i}(t)$, $\hat e_{i|j}(t)$ does not depend on $e_j(t)$ and there is no increase in  variance of $\hat G_i$, cf. Remark~\ref{rem:theta_var_remark_explanation}. 
\end{rem}

\subsection{A graphical representation of Corollary~\ref{cor:variance_increase}}

Following the notation in Bayesian Networks \citep{koski2012review}, Conditions 1) and 2) in Corollary~\ref{cor:variance_increase} can be interpreted graphically. 
	Each module is represented by a vertex in a weighted directed acyclic graph $\Gc$. Let the vertices be ordered by model order, \ie let the first vertex correspond to $\hat G_1$.
	Under Assumption~\ref{ass:input_noise}, with the additional assumption that module $i$ is the first module with order $n_i$, let there be an edge, denoted by $j \to i$, from vertex $j$ to $i$, $j < i$, if $e_i(t)$ is not orthogonal to $e_j(t)$ conditioned on $e_{j \backslash i}(t)$. Notice that this is equivalent to $\gamma_{ij} \neq 0$. 
	Let the weight of the edge be $\gamma_{ij}$ and define the parents of vertex $i$ to be all nodes with a link to vertex $i$, i.e., $pa_{\Gc} (i) \defeq \{j: j \to i \}$. 
	Then, \eqref{eq:cor_var_theta_k}, together with Lemma~\ref{lem:variance_non_estimable_part_of_noise}, shows that only outputs corresponding to parents of node $i$ affect the asymptotic variance. 
Indeed, a vertex without parents has variance 
\begin{IEEEeqnarray*}{rCl}  \label{eq:graph_no_parents}
\asvar{\hat G_i } &=& \frac{\lambda_i}{\sigma^2}  \sum_{k=1}^{n_i}|\Bc_k(e^{j\omega_0})|^2 \!,
\yesnumber
\end{IEEEeqnarray*} 
which corresponds to \eqref{eq:ascov_G_i_hat} with 
\begin{IEEEeqnarray*}{rCcCl} 
\lambda_{i|0} & = & \dotsc & = &  \lambda_{i|i-1} = \lambda_i. 
\end{IEEEeqnarray*} 
Thus, $\asvar{ \hat G_i } $ is independent of the model order of the other modules. 

\begin{figure}[!ht]
    \centering    
	\includegraphics[scale=1]{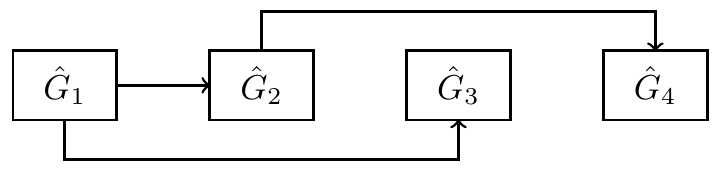}
	\caption{Graphical representation of Conditions 1) and 2) in Corollary~\ref{cor:variance_increase} for the Cholesky factor given in \eqref{eq:chol_factor_example}.}
	 \label{fig:noise_graph}
\end{figure}

As an example, consider four systems with the lower Cholesky factor of the covariance of $e(t)$ given by:
\begin{IEEEeqnarray*}{rCl}  \label{eq:chol_factor_example}
\Lambda &=& 
\begin{bmatrix}
1 			&	 0.4 		 &	 	0.2	 		& 		0  \\
0.4	 		&	 1.16		 &		0.08			&	 	0.3  \\
0.2			&	0.08			 &		1.04			& 		0  	\\
0 			&	0.3			 &	 	0			& 		1.09  
\end{bmatrix} \\
&=&
\underbrace{
\begin{bmatrix}
1		 	&	 0	 		 &	 	0	 		& 		0  \\
0.4		 	&	 1			 &		0			&	 	0  \\
0.2			&	0			 &		1			& 		0  	\\
0 			&	 0.3			 &	 	0			& 		1  
\end{bmatrix}}_{\Lambda_{CH}}
\begin{bmatrix}
1		 	&	 0	 		 &	 	0	 		& 		0  \\
0.4		 	&	 1			 &		0			&	 	0  \\
0.2			&	0			 &		1			& 		0  	\\
0 			&	 0.3			 &	 	0			& 		1  
\end{bmatrix}^T
\yesnumber
\end{IEEEeqnarray*}
If the model orders are distinct, the corresponding graph is given in Figure~\ref{fig:noise_graph}, where one can see that $\asvar \hat G_4$ depends on $y_2$ (and on $y_4$ of course), but depends neither on $y_3$ nor $y_1$ since $\gamma_{43} = \gamma_{41} = 0$,  $\asvar \hat G_3$ depends on $y_1$, but not on $y_2$ since $\gamma_{32} = 0$ and $\asvar \hat G_2$ depends on $y_1$, while the variance of  $\hat G_1$ is given by \eqref{eq:graph_no_parents}. 
If $n_2 = n_4$, the first condition of Corollary~\ref{cor:variance_increase} is not satisfied and we have to cut the edge between $\hat{G}_4$ and $\hat{G}_2$. Similarly, if $n_1 = n_2$, we have to cut the edge between $\hat{G}_2$ and $\hat{G}_1$, and if additionally $n_1 = n_2 = n_3$, we have to cut the edge between $\hat{G}_3$ and $\hat{G}_1$. 

\section{Connection between MISO and SIMO}
\label{sec:connection-MISO}
There is a strong connection between the results presented here and those regarding MISO systems presented in~\citet{Ramazi20141675}. We briefly restate the problem formulation and some results from~\citet{Ramazi20141675} to show the connection. The MISO data generating system is in some sense the dual of the SIMO case. With $m$ spatially correlated inputs and one output, a MISO system is described by 
\begin{IEEEeqnarray*}{rCl}
y(t) =
\begin{bmatrix}
G_{1}(q)  &
G_{2}(q) 	&
\ldots		&
G_{m}(q) 	
\end{bmatrix}  
\begin{bmatrix}
u_{1}(q)  \\
u_{2}(q)	  \\
\vdots	  \\
u_{m}(q) 	
\end{bmatrix}
+ e(t).
\end{IEEEeqnarray*}
The input sequence $\{u(t)\}$ is zero mean and temporally white, but may be correlated in the spatial domain,
\begin{IEEEeqnarray*}{rCl}
\expect{u(t)} &=& 0\\
\expect{u(t)u(s)^T} &=& \delta_{t-s} \Sigma_u,
\end{IEEEeqnarray*}
for some positive definite matrix covariance matrix $\Sigma_u = \Sigma_{CH} \Sigma_{CH} ^T$, where $\Sigma_{CH}$ is the lower triangular Cholesky factor of $\Sigma$. 
The noise $e(t)$ is zero mean and has variance $\lambda$.
The asymptotic covariance of the estimated parameters can be expressed using \eqref{eq:inp_psi} with 
\begin{IEEEeqnarray*}{rCl} \label{eq:MISO_psi}
\Psi = \Psi^{\mathrm{MISO}} \defeq \tilde \Psi  \Sigma_{CH}.
\yesnumber
\end{IEEEeqnarray*} 
We make the convention that $\sum_{k=k_1}^{k_2} x_k = 0$ whenever $k_1 > k_2$.

\begin{thm}[Theorem 4 in~\citet{Ramazi20141675}]
\label{thm:MISO_variance_G_i}
Under Assumption~\ref{ass:input_noise}, but with $n_1 \ge n_2 \ge \dotsc, \ge n_m$,  
for any frequency $\omega_0$ it holds that 
\begin{IEEEeqnarray*}{rCl}
&\asvar \hat  G_i = \sum_{j=i}^m \frac{\lambda}{\sigma^2_{i|j} } \sum_{k=n_{j+1}+1}^{n_j} |\Bc_k(e^{j\omega_0})|^2 \! ,
\IEEEeqnarraynumspace \yesnumber \label{eq:MISO_original_modulevariance}
\end{IEEEeqnarray*} 
where $n_{m+1} \defeq 0$ and $\sigma^2_{i|j}$ is the variance of the non-estimable part of $u_i(t)$ given $u_{j \backslash i}(t)$.
\end{thm}
\begin{cor}[Corollary 6 in~\citet{Ramazi20141675}]
\label{cor:MISO_variance_increase}
Under Assumption~\ref{ass:input_noise}, but with $n_1 \ge n_2 \ge \dotsc, \ge n_m$. 
Suppose that the order of block $j$ is increased from $n_j$ to $n_{j+1}$. Then there is an increase in the asymptotic variance of $\hat G_i$ if and only if all the following conditions hold:
\begin{enumerate}
\item $n_j< n_i$,
\item $u_i(t)$ is not orthogonal to $u_j(t)$ conditioned on $u_{j \backslash i}(t)$,
\item $ |\Bc_{n_{j+1}}(e^{j\omega_0})|^2 \neq 0$.
\end{enumerate}
\end{cor}
\begin{rem}
The similarities between Corollary~\ref{cor:variance_G_i} and Theorem~\ref{thm:MISO_variance_G_i}, and between Corollary~\ref{cor:variance_increase} and Corollary~\ref{cor:MISO_variance_increase} are striking. In both cases it is the non-estimable part of the input and noise, respectively, along with the estimated basis functions that are the key determinants for the resulting accuracy.
Just as in Corollary~\ref{cor:variance_G_i}, Theorem~\ref{thm:MISO_variance_G_i} can be expressed with respect to the basis functions: 
\begin{IEEEeqnarray*}{rCl}
\asvar \hat G_i  &=& \sum_{k=1}^{n_i} \asvar \hat \theta_{i,k} \: |\Bc_k(e^{j\omega_0})|^2.
\IEEEeqnarraynumspace \yesnumber \label{eq:MISO_modulevariance}
\end{IEEEeqnarray*} 
However, now 
\begin{IEEEeqnarray*}{rCl}
 \asvar \hat \theta_{i,k} &=& \frac{\lambda}{\sigma^2_{i|\chi_k} }
 \IEEEeqnarraynumspace \yesnumber \label{eq:MISO_theta_variance}
\end{IEEEeqnarray*} 
where $\sigma^2_{i|l}$ is determined by the correlation structure of the inputs $u_i(t)$ to the systems $G_i(q,\theta_i)$ that \emph{do} share basis function $\Bc_k(q)$ ($i = 1,\dotsc, \chi_k$). Note that in the SIMO case we had 
\begin{IEEEeqnarray*}{rCl}
 \asvar \hat \theta_{i,k} &=& \frac{\lambda_{i|\chi_k}}{\sigma^2 }
\end{IEEEeqnarray*} 
where $\lambda_{i|\chi_k}$ is determined by the correlation structure of the noise sources $e_i(t)$ affecting systems $G_i(q,\theta_i)$ that \emph{do not} share basis function $\Bc_k(q)$ ($i = 1,\dotsc, \chi_k$). Note that \eqref{eq:MISO_original_modulevariance} found in \citet{Ramazi20141675} does not draw the connection to the variance of the parameters. This is made explicit in the alternate expressions \eqref{eq:MISO_theta_variance} and \eqref{eq:MISO_modulevariance}.
\end{rem}

The correlation between parameters related to the same basis functions is not explored in \citet{Ramazi20141675}. 
In fact, it is possible to follow the same line of reasoning leading to Theorem~\ref{thm:cov_diag} and arrive at the counter-part for MISO systems. Let the first $\chi_k$ systems contain basis function $k$, so 
\begin{IEEEeqnarray*}{rCl}
\asvar \hat {\bar{\theta}}_k^{MISO}   &=& \lambda \: \Sigma_{1:\chi_k}^{-1}
\end{IEEEeqnarray*} 
where $\Sigma_{1:\chi_k}$ denotes the covariance matrix of the first $\chi_k$ inputs. Hence 
\begin{IEEEeqnarray*}{rCl} \label{eq:thm_MISO_ascov_G}
\ascov \hat {G} &=& 
\lambda 
\sum_{k=1}^{n_1} 
\begin{bmatrix}
\Sigma_{1:\chi_k}^{-1} 	& 	\bf{0} \\
\bf{0}   &  	{\bf{0}}_{m-\chi_k}
\end{bmatrix}
|\Bc_{k}(e^{j\omega_0})|^2
,
\IEEEeqnarraynumspace
\end{IEEEeqnarray*}
and
\begin{IEEEeqnarray*}{rCl} \label{eq:cor_MISO_diag}
 	\ascov \hat{ \bar{\theta}}^{MISO} 
 	&=&  \lambda  \: \diag( \Sigma_{1:\chi_1}^{-1}, \Sigma_{1:\chi_2}^{-1}, \dotsc,  \Sigma_{1:\chi_{n_m}}^{-1} ).
 	\IEEEeqnarraynumspace
 	\yesnumber
\end{IEEEeqnarray*}
Note that, while the correlation between the noise sources is beneficial, the correlation in the input is detrimental for the estimation accuracy.
Intuitively, if we use the same input to parallel systems, and only observe the sum of the outputs,  there is no way to determine the contribution from the individual systems. On the other hand, as observed in the example in Section~\ref{subsec:example}, if the noise is correlated, we can construct equations with reduced noise and improve the accuracy of our estimates.

This difference may also be understood from the structure of $\Psi$, which through \eqref{eq:inp_psi} determines the variance properties of any estimate. 
Consider a single SISO system $G_1$ as the basic case.
For the SIMO structure considered in this paper, as noted before, $\Psi^{\mathrm{SIMO}}$ of \eqref{eq:SIMO_psi} is block upper triangular with $m$ columns (the number of outputs), while $\Psi^{\mathrm{MISO}}$ is block lower triangular with as many columns as inputs. $\Psi^{\mathrm{MISO}}$ is block lower triangular since $\tilde \Psi$ is block diagonal and $\Sigma_{CH}$ is lower triangular in \eqref{eq:MISO_psi}. Adding an output $y_j$ to the SIMO structure corresponds to extending $\Psi^{\mathrm{SIMO}}$ with one column (and $n_j$ rows):
\begin{IEEEeqnarray*}{rCl}
\Psi^{\mathrm{SIMO}}_e = \begin{bmatrix}
\Psi^{\mathrm{SIMO}} & \star \\
0 &	\star 
\end{bmatrix},
\IEEEeqnarraynumspace \yesnumber \label{eq:SIMO_added_column}
\end{IEEEeqnarray*}
where the zero comes from that $\Psi^{\mathrm{SIMO}}_e$ also is block upper triangular.
Since $\Psi^{\mathrm{MISO}}$ is block lower triangular, adding an input $u_j$ to the MISO structure extends $\Psi^{\mathrm{MISO}}$ with $n_j$ rows (and one column):
\begin{IEEEeqnarray*}{rCl}
\Psi^{\mathrm{MISO}}_e = \begin{bmatrix}
\Psi^{\mathrm{MISO}} & 0 \\
\star &	\star 
\end{bmatrix},
\IEEEeqnarraynumspace \yesnumber \label{eq:MISO_added_rows}
\end{IEEEeqnarray*}
where $\star$ denotes the added column and added row respectively. 
Addition of columns to $\Psi$ decreases the variance of $G_1$, while addition of rows increases the variance. First, a short motivation of this will be given. Second, we will discuss the implication for the variance analysis. 

Addition of one more column to $\Psi$ in \eqref{eq:SIMO_added_column} decreases the variance of $G_1$.
With $\Psi = \begin{bmatrix}
\psi_1 & \dotsc & \psi_m
\end{bmatrix}$, $\inp{\Psi,\Psi} = \sum_k^m \inp{\psi_k,\psi_k}$, where $\inp{\psi_k,\psi_k} \ge 0$ for every $k$. The variance of the parameter estimate $\hat \theta_N$  decreases with the addition of a column, since
\begin{IEEEeqnarray*}{rCl}
 \inp{\Psi_e,\Psi_e}^{-1} &\le & \left[ \inp{\psi_{m+1},\psi_{m+1}} +  \inp{\Psi,\Psi} \right]^{-1}.
\end{IEEEeqnarray*}
On the other hand, addition of rows leads to an increase in variance of $\hat G_1$, e.g., consider \eqref{eq:ascov_as_sum_of_basis_functions} in Lemma \ref{lem:ascov_as_sum_of_basis_functions}, 
\begin{IEEEeqnarray*}{rCl}
\ascov G_1(\hat \theta_N) = L^T \sum_{k=1}^r \Bc_k^\Sc(z_o)^* \Bc_k^\Sc(z_o) \:  L, \IEEEeqnarraynumspace
\end{IEEEeqnarray*}
where 
$L = \Sigma_{CH} ^{-T}
\begin{bmatrix}
1 & 0 & \dotsc & 0
\end{bmatrix}^T $
 for any number of inputs, and $\{\Bc_k^\Sc \}_{k=1}^r$ is a basis for the linear span of the rows of $\Psi^{\mathrm{MISO}}$.
As can be seen from \eqref{eq:MISO_added_rows}, the first rows of $\Psi^{\mathrm{MISO}}_e$ are the same as for $\Psi^{\mathrm{MISO}}$ and the first $r$ basis functions can therefore be taken the same (with a zero in the last column). 
To accommodate for the extra rows, $n_e$ extra basis functions $\{\Bc_k^\Sc \}_{k=r+1}^{r+n_e}$ are needed. 
Thus, $\{\Bc_k^\Sc \}_{k=1}^{r+n_e}$ is a basis for the linear span of $\Psi^{\mathrm{MISO}}_e$.
We see that the variance of $G_1(\hat \theta_N^e)$ is larger than $G_1(\hat \theta_N)$ since
\begin{IEEEeqnarray*}{rCl}
\ascov G_1(\hat \theta_N^e) &= & \ascov G_1(\hat \theta_N) \\
&&  + 
L^T \sum_{k=r+1}^{r+n_e} \Bc_k^\Sc(z_o)^* \Bc_k^\Sc(z_o) \: L,
\IEEEeqnarraynumspace
\end{IEEEeqnarray*}
and $L^T \Bc_k^\Sc(z_o)^*\Bc_k^\Sc(z_o) L \ge 0 $ is positive semidefinite for every $k$.

Every additional input of the MISO system corresponds to addition of rows to $\Psi$.
The increase is strictly positive provided that the explicit conditions in Corollary~\ref{cor:MISO_variance_increase} hold. 

Every additional output of the SIMO system corresponds to the addition of one more column to $\Psi$. However, the benefit of the additional columns is reduced by the additional rows arising from the additional parameters that need to be estimated, cf. Corollary~\ref{cor:variance_increase} and the preceding discussion. When the number of additional parameters has reached $n_1$ or if $e_1(t)$ is orthogonal to $e_j(t)$ conditioned on $u_{j \backslash 1}(t)$ the benefit vanishes completely. The following examples clarify this point.

\begin{exmp} \label{ex:fir_psi_simo}

We consider the same example as in Section~\ref{subsec:example} for three cases of model orders of the second model, $n_2 = 0,1,2$. 
These cases correspond to $\Psi$ given by the first $n_2+1$ rows of 
\begin{IEEEeqnarray*}{rCl}
\Psi^{SIMO}_e(q) = \begin{bmatrix}
q^{-1} & - \sqrt{1-\beta^2}/\beta q^{-1}  \\
0 & 1/\beta q^{-1}	\\
0 & 1/\beta q^{-2}	\\
\end{bmatrix},
\end{IEEEeqnarray*}
respectively. 
When only $y_1$ is used ($\Psi = q^{-1}$) :
\begin{IEEEeqnarray*}{rCcCcCl}
\asvar \hat \theta_{1,1} &=& \inp{\Psi, \Psi }^{-1} &=& 1. \IEEEeqnarraynumspace
\end{IEEEeqnarray*}
When $n_2=0$, the second measurement gives a benefit determined by how strong the correlation is between the two noise sources:
\begin{IEEEeqnarray*}{rCcCcCl}
\asvar \hat \theta_{1,1} &=& \inp{\Psi, \Psi }^{-1} &=& (1+ (1-\beta^2)/\beta^2)^{-1} &=& \beta^2. \IEEEeqnarraynumspace
\end{IEEEeqnarray*}
However, already if we have to estimate one parameter in $G_2$ the benefit vanishes completely, i.e., for $n_2=1$:
\begin{IEEEeqnarray*}{rCcCcCl}
\ascov \hat \theta &=& \inp{\Psi, \Psi }^{-1} &=& 
\begin{bmatrix}
1 & \sqrt{1 - \beta^2}  \\
\sqrt{1 - \beta^2} & 1
\end{bmatrix}.
 \IEEEeqnarraynumspace
\end{IEEEeqnarray*}
The third case, $n_2=2$, corresponds to the example in Section~\ref{subsec:example}, which shows that the first measurement $y_1$ improves the estimate of $\theta_{2,2}$ (compared to only estimating $\hat{G}_2$ using $y_2$):
\begin{IEEEeqnarray*}{rCcCcCl}
\asvar \hat \theta_{2,2} &=& 
\lambda_{2|1} = \beta^2.
 \IEEEeqnarraynumspace
\end{IEEEeqnarray*}
\end{exmp}  

\begin{exmp} \label{ex:fir_psi_miso}
We consider the corresponding MISO system with unit variance noise $e(t)$ and $u(t)$ instead having the same 
spectral factor
\begin{IEEEeqnarray*}{rCl}
\Sigma_{CH} &=& 
\begin{bmatrix}
1 & 0 \\ \sqrt{1-\beta^2} & \beta
\end{bmatrix}.
\end{IEEEeqnarray*}
for $\beta \in (0,1]$.
We now study the impact of the second input $u_2(t)$ when  
\begin{IEEEeqnarray*}{rCl}
G_1(q) = \theta_{1,1}q^{-1}+\theta_{1,2}q^{-2}, \quad G_2(q) = \sum_{k=1}^{n_2} \theta_{2,k}q^{-k} \IEEEeqnarraynumspace
\end{IEEEeqnarray*}
 and $n_2 = 0,1$. 
 These two cases correspond to $\Psi$ given by the first $n_2+2$ rows of 
\begin{IEEEeqnarray*}{rCl}
\Psi^{MISO}_e(q) = \begin{bmatrix}
q^{-1} & 0  \\
q^{-2} & 0	\\
\sqrt{1-\beta^2}q^{-1} & \beta q^{-1}	\\
\end{bmatrix},
\end{IEEEeqnarray*}
respectively. 
When only $u_1$ is used or $G_2$ is known ($\Psi = \begin{bmatrix}
q^{-1} & q^{-2}
\end{bmatrix}^T$):
\begin{IEEEeqnarray*}{rCcCcCl}
\ascov \hat \theta_{1} &=& \inp{\Psi, \Psi }^{-1} &=& I. \IEEEeqnarraynumspace
\end{IEEEeqnarray*}
When $n_2=1$, the variance of $\hat \theta_{1,1}$ is increased depending on the correlation between the two inputs:
\begin{IEEEeqnarray*}{rCl}
\asvar \hat \theta &=& 
\frac{1}{\beta^{2}}
\begin{bmatrix}
1 & 0  & -\sqrt{1 - \beta^2} \\
0 & \beta^{2} & 0 \\
-\sqrt{1 - \beta^2} & 0 & 1
\end{bmatrix}.
\IEEEeqnarraynumspace
\end{IEEEeqnarray*}
Also notice that the asymptotic covariance of $\begin{bmatrix}
\hat \theta_{1,1} & \hat \theta_{2,1}
\end{bmatrix}^T$ is given by $\Sigma^{-1}$, the inverse of the covariance matrix of $u(t)$ and that $\asvar \: \hat \theta_{2,2} = \sigma_1^{-1}$. As $\beta$ goes to zero the variance of  $\begin{bmatrix}
\hat \theta_{1,1} & \hat \theta_{2,1}
\end{bmatrix}^T$ increases and at $\beta = 0$ the two inputs are perfectly correlated and we loose identifiability.
\end{exmp}  

\section{Effect of input spectrum}
\label{sec:input spectrum}
In this section we will see how a non white input spectrum changes the
results of Section~\ref{sec:main}.
Using white noise filtered through an AR-filter as input, we may use the developed results to show
where in the frequency range the benefit of the correlation structure is focused. 
An alternative approach, when a non-white input spectrum is used, is to change the basis functions as discussed in Remark~\ref{rem:model_structure}. 
However, the effect of the input filter is internalized in the basis functions and it is hard to distinguish the effect of the input filter. 
We will instead use FIR basis functions for the SIMO system, which are not orthogonal with respect to the inner product induced by the input spectrum, cf. Remark~\ref{rem:model_structure}.
We let the input filter be given by 
\begin{IEEEeqnarray*}{rCl}  \label{eq:input_filter}
u(t) &=&  \frac{1}{A(q)} w(t)
\yesnumber
\end{IEEEeqnarray*}
where $w(t)$ is a white noise sequence with variance $\sigma_w^2$ and the order $n_a$ of $A$ is less than the order of $G_1$, \ie $n_a \le n_1$. 
In this case, following the derivations of Theorem~\ref{thm:cov_diag},
it can be shown that 
\begin{IEEEeqnarray*}{rCl}
\label{eq:ascov_G_i_hat_ARX}
\asvar \hat G_i  &=& \sum_{k=1}^{n_i}  \asvar \hat \theta_{i,k} |\Bc_k(e^{j\omega_0})|^2
\yesnumber
\end{IEEEeqnarray*} 
where
\begin{IEEEeqnarray*}{rCl}
\asvar \hat \theta_{i,k} &=& \frac{\lambda_{i|\chi_k-1}}{ \Phi_u(\omega_0) }
\end{IEEEeqnarray*} 
and the basis functions $\Bc_k$ have changed due to the input filter. 
The solutions boil down to finding explicit basis functions $\Bc_k$ for the
 well known case
\citep{Ninness&Gustafsson:97} when
\begin{align*}
 \Span \left \{\frac{\Gamma_n}{A(q)} \right\}=\Span \left\{\frac{q^{-1}}{A(q)},\frac{q^{-2}}{A(q)},\ldots,\frac{q^{-n}}{A(q)}\right\} \nonumber
\end{align*}
where $A(q)=\prod_{k=1}^{n_a}(1-\xi_kq^{-1})$, $|\xi_k|<1$ for
some set of specified poles $\{\xi_1,\ldots,\xi_{n_a}\}$ and where
$n\geq n_a$. Then, it holds \citep{Ninness&Gustafsson:97} that
\begin{align*}
  \Span \left \{\frac{\Gamma_n}{A(q)} \right\} =\Span\left\{\mathcal{B}_1,\ldots, \mathcal{B}_n\right\}
\end{align*}
where $\{\mathcal{B}_k\}$ are the Takenaka-Malmquist functions
given by
\begin{align*}
\label{Bk}
\mathcal{B}_k(q)&:=\frac{\sqrt{1-|\xi_k|^2}}{q-\xi_k}\; 
\phi_{k-1}(q),\; k=1,\ldots, n  \\ 
\phi_k(q)&:=\prod_{i=1}^{k}\frac{1-\overline{\xi_i}q}{q-\xi_i},\;\;\phi_0(q):=1
\yesnumber
\end{align*}
and with $\xi_k=0$ for $k=n_a+1,\ldots,n$. 
We summarize the result in the following theorem:

\begin{thm}
Let the same assumptions as in Theorem~\ref{thm:cov_diag} hold. 
Additionally the input $u(t)$ is generated by an AR-filter as in \eqref{eq:input_filter}. Then for any frequency $\omega_0$ it holds that
\begin{IEEEeqnarray*}{rCl}
\asvar \hat G_i  &=& \frac{1}{\Phi_u(\omega_0)} \Bigg( \lambda_{i} \sum_{k=1}^{n_a} \frac{1-|\xi_k|^2}{|e^{j\omega_0}-\xi_k|^2}  +  \lambda_{i} (n_1-n_a) \\
&&+ 
 \sum_{j=2}^i \lambda_{i|j-1} (n_j-n_{j-1}) \Bigg)
\yesnumber 
\label{eq:var-non-white-input}
\end{IEEEeqnarray*} 
\end{thm}

\begin{pf}
The proof follows from \eqref{eq:ascov_G_i_hat_ARX}
with the basis functions given by the Takenaka-Malmquist functions and using that $\phi_{k-1}(q)$ is all-pass, and for $k > n_a$, also  $\Bc_k(q)$ is all-pass. This means that  $ |\Bc_k(\eiw)|^2 = 1$ for all $k>n_a$. $\hfill \blacksquare$
\end{pf}

\begin{rem}
The second sum in \eqref{eq:var-non-white-input} is where the benefit from the correlation structure of the noise at the other sensors enters through $\lambda_{i|j-1}$.
This contribution is weighted by $1/\Phi_u(\omega)$. The benefit thus enters mainly where $\Phi_u(\omega)$ is small. The first sum gives a variance contribution that is less focused around frequencies close to the poles ($|e^{j\omega_0}-\xi_k|^2$ is canceled by $\Phi_u(\omega)$). This contribution is not reduced by correlation between noise sources. 
Shaping the input thus gives a weaker effect on the asymptotic variance than what is suggested by the asymptotic in model order result \eqref{eq:asym-model-order}, and what would be the case if there would be no correlation between noise sources (replacing $\lambda_{i|j-1}$ by $\lambda_{i}$ in \eqref{eq:var-non-white-input}).
\end{rem}

\begin{figure}[!ht]
    \centering    
	\input{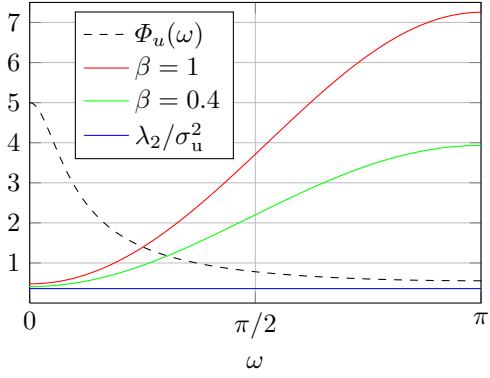}
	\caption{Asymptotic variance of $\hat G_3(e^{j\omega}, \hat \theta_2)$ for $\beta = 1$ and $\beta = 0.4$. 
	Also shown is $\lambda_2/\sigma^2$, the first term of $\asvar \hat G_3(e^{j\omega}, \hat \theta_2)$ in  \eqref{eq:example_var-non-white-input}.}
	 \label{fig:input}
\end{figure}

For the example in Section~\ref{subsec:example} for filtered input with $n_1 = 2$ , $n_2 = 3$ and $n_a = 1$,  
\eqref{eq:var-non-white-input} simplifies to
\begin{IEEEeqnarray*}{rCl}
\asvar \hat G_2  &=& 
\frac{\lambda_2}{\sigma^2}
+ 
\frac{\lambda_2}{\Phi_u(\omega_0)} 
+ 
\frac{\lambda_{2|1}}{\Phi_u(\omega_0)} .
\yesnumber \IEEEeqnarraynumspace
\label{eq:example_var-non-white-input}
\end{IEEEeqnarray*} 
In Figure~\ref{fig:input} the variance of $\hat G_2$ for an input filtered with $A(q) = 1-0.8 q^{-1}$ is presented. The filter is of low-pass type and thus gives high input power at low frequencies which results in low variance at those frequencies. Correlation between noise sources decreases the variance mainly where $\Phi_u(\omega)$ is small, \ie at higher frequencies.

\section{Optimal correlation structure}
\label{sec:optimal_correlation_structure}
In this section we characterize the optimal correlation structure, in order to minimize the total variance. 
In the previous section, we have seen that not estimating a basis function $k$ leads to a reduction in variance of the other parameters related to basis function $k$. In this section, we will find the optimal correlation structure when the parameterization of the system is such that $n_1+1 = n_2 = \ldots = n_m$, \ie the first module has one parameter less than the others.
Let 
$\tilde \theta \defeq [\theta_{2,n_2},\,\theta_{3,n_2},\,\ldots,\,\theta_{m,n_2}]^T,$
 i.e., the sub vector of parameters related to basis function $\Bc_{n_2}(q)$. 
Assume the SIMO structure given by \eqref{eq:model} and let the input be white noise with unit variance. Recalling Theorem \ref{thm:cov_diag}, the covariance matrix of $\hat{\theta}$ is given by $\Lambda_{2:m|1}$. In particular, the variance of the entries of $\tilde \theta$ is
\begin{IEEEeqnarray*}{rCl} \label{eq:variance_one_less}
\asvar \hat \theta_{k,n_2} & =& \frac{\lambda_{k|1}}{\sigma^2}  \,,  \quad k = 2,\dotsc, m.
\yesnumber
\end{IEEEeqnarray*}
As before, $\lambda_{k|1}$ is the non-estimable part of $e_k(t)$ given $e_1(t)$.

%

Recalling Theorem \ref{thm:cov_diag}, the covariance matrix of $\tilde{\theta}$ is given by $\frac{1}{\sigma^2} \Lambda_{2:m|2}$. The total variance of $\hat{\tilde{\theta}}$ is defined as
\begin{IEEEeqnarray*}{rCl}
\Tvar \tilde \theta & \defeq &  \sum_{i=2}^{m} \asvar \hat {\theta}_{i,n_2}
= \Tr \frac{1}{\sigma^2}		\Lambda_{2:m|1}
 \,.
\end{IEEEeqnarray*}
We are interested in understanding how the correlation of $e_1(t)$ with $e_2(t),\,\ldots,\,e_{m}(t)$, i.e., $\Lambda_{12}\,\,\ldots,\,\Lambda_{1m}$,  should be in order to obtain the minimum value of the above total variance. This problem can be expressed as follows:
\begin{IEEEeqnarray*}{cCc} \label{eq:min_total_variance}
\minimize_{\Lambda_{12}\,\,\ldots,\,\Lambda_{1m}} \quad & \Tr \Lambda_{2:m|1} \\
\mbox{s.t.} &    \Lambda \ge 0,
\yesnumber
\end{IEEEeqnarray*}
where the constraint on $\Lambda$ implies that not all choices of the entries $\Lambda_{1i}$ are allowed. Directly characterizing the noise structure using this formulation of the problem appears to be hard. Therefore, it turns out convenient to introduce an upper triangular Cholesky factorization of $\Lambda$, namely define $B$ \textit{upper triangular} such that $\Lambda = BB^T$. Note that
\begin{enumerate}
\item the rows of $B$, $b_i^T$, $i=1,\,\ldots,\,m$, are such that $\|b_i\|^2 = \lambda_i$;
\item $\mathbf{E}\{e_1 e_i\} = \Lambda_{1i} = b_1^Tb_i$;
\item there always exists an orthonormal matrix $Q$ such that $B = \Lambda_{CH}Q$, where $\Lambda_{CH}$ is the previously defined lower triangular Cholesky factor.
\end{enumerate}
\begin{lem} \label{lem:proof_lemma_inversion}
Let
\begin{IEEEeqnarray*}{rCl}
B &=& \begin{bmatrix} \eta & p^T \\ 0 & M \end{bmatrix}, \, M \in \mathbb{R}^{m-1 \times m-1} \! ,\, p \in \mathbb{R}^{m-1 \times 1} \! , \, n \in \mathbb{R} \,;
\end{IEEEeqnarray*}
then
\begin{IEEEeqnarray*}{rCl} \label{eq:one_less_parameter}
\Lambda_{2:m|1} = M(I - \frac{1}{\lambda_1})pp^TM^T  \,.
\yesnumber
\end{IEEEeqnarray*}
\end{lem}
\begin{pf}
See Appendix \ref{sec:proof_lemma_inversion}. $\hfill \blacksquare$
\end{pf}
Using the previous lemma we reformulate \eqref{eq:min_total_variance}; keeping $M$ fixed and letting $p$ vary, we have
\begin{IEEEeqnarray*}{cCc} \label{eq:min_total_variance_th}
\maximize_{b_1} \quad & \Tr \; \frac{1}{\lambda_1}Mpp^TM^T \\
\mbox{s.t.} & \|b_1\|^2 = \lambda_1 \,,
\yesnumber
\end{IEEEeqnarray*}
with $b_1^T = [\eta \quad p^T]$. Note that the constraint $\Lambda \geq 0$ is automatically satisfied.
Let us define $v_1,\,\ldots,\,v_{m-1}$ as the right-singular vectors of $M$, namely the columns of the matrix $V$ in the singular value decomposition (SVD) $M = USV^T$. The following result provides the structure of $B$ that solves \eqref{eq:min_total_variance}.
\begin{thm} \label{th:optimal_correlation}
Let the input be white noise. Let $n_1+1 = n_2 = \ldots = n_m$. Then \eqref{eq:min_total_variance} is solved  by an upper triangular Cholesky factor $B$ such that its first row is
\begin{IEEEeqnarray*}{rCl} \label{eq:optimal_l}
b_1^{*T} &=& \lambda_1[0 \quad v_1^T] \,.
\yesnumber
\end{IEEEeqnarray*}
\end{thm}
\begin{pf}
Observe the following facts:
\begin{enumerate}
\item $\Tr \; \frac{1}{\lambda_1}Mpp^TM^T = \frac{1}{\lambda_1}p^T M^T  M p = \frac{1}{\lambda_1}\|Mp\|^2$;
\item since $b_1^T = [\eta \quad p^T]$, it is clear that a candidate solution to \eqref{eq:min_total_variance_th} is of the form $b_1^{*T} = [0 \quad p^{*T}]$.
\end{enumerate}
It follows that Problem \eqref{eq:min_total_variance_th} can be written as
\begin{IEEEeqnarray*}{cCc} \label{eq:min_total_variance2}
\maximize_{b_1} \quad &  \|Mp\|^2 \\
\mbox{s.t.} & \|p\|^2 = \lambda_1 \,,
\yesnumber
\end{IEEEeqnarray*}
whose solution is known to be (a rescaling of) the first right singular vector of $M$, namely $v_1$. Hence $b_1^{*T} = \lambda_1[0 \quad v_1^T]$. $\hfill \blacksquare$
\end{pf}
\begin{rem}
As has been pointed out in Section \ref{sec:problem_statement}, $\Lambda$ is required to be positive definite. Thus, the ideal solution provided by Theorem \ref{th:optimal_correlation} is not applicable in practice, where one should expect that $\eta$, the first entry of $b_1$, is always nonzero. In this case, the result of Theorem \ref{th:optimal_correlation} can be easily adapted, leading to $b_1^{*T} = \left[\eta \quad \sqrt{1-\frac{\eta^2}{\lambda_1}} v_1^T \right]$.
\end{rem}

\section{Numerical examples} \label{sec:num_exp}

In this section, we illustrate the results derived in the previous sections through three sets of numerical Monte Carlo simulations.

\subsection{Effect of model order} \label{sec:example_model_order}

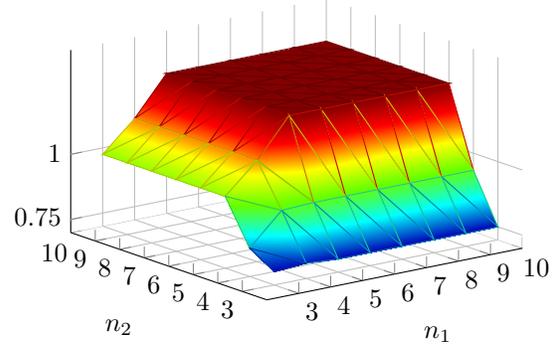
\begin{figure}
    \centering
%
%
\begin{tikzpicture}

\begin{axis}[%
width=\figurewidth,
height=\figureheight,
view={-37.5}{25},
scale only axis,
xmin=2,
xmax=10,
xtick={ 3,  4,  5,  6,  7,  8,  9, 10},
xlabel={$n_1$},
xmajorgrids,
ymin=2,
ymax=10,
ytick={ 3,  4,  5,  6,  7,  8,  9, 10},
ylabel={$n_2$},
ymajorgrids,
zmin=0.7,
zmax=1.4,
ztick={ 0.5, 0.75,   1},
zlabel={},
zmajorgrids,
axis x line*=bottom,
axis y line*=left,
axis z line*=left
]

\addplot3[%
surf,
shader=faceted interp,
colormap/bluered,
mesh/rows=8]
table[row sep=crcr,header=false] {
3 3 0.750485648194549\\
3 4 0.808969914367952\\
3 5 0.975462220011134\\
3 6 0.975462220011154\\
3 7 0.975462220011136\\
3 8 0.975462220011138\\
3 9 0.97546222001115\\
3 10 0.975462220011118\\
4 3 0.758484706037414\\
4 4 0.928714799089149\\
4 5 1.08765616245725\\
4 6 1.08765616245724\\
4 7 1.08765616245726\\
4 8 1.08765616245723\\
4 9 1.08765616245724\\
4 10 1.08765616245724\\
5 3 0.749638072622369\\
5 4 0.93454195961623\\
5 5 1.23609271356532\\
5 6 1.23609271356531\\
5 7 1.23609271356531\\
5 8 1.23609271356531\\
5 9 1.23609271356531\\
5 10 1.23609271356533\\
6 3 0.74963807262238\\
6 4 0.934541959616206\\
6 5 1.23609271356531\\
6 6 1.2360927135653\\
6 7 1.23609271356531\\
6 8 1.2360927135653\\
6 9 1.23609271356529\\
6 10 1.23609271356531\\
7 3 0.749638072622381\\
7 4 0.934541959616228\\
7 5 1.23609271356525\\
7 6 1.23609271356527\\
7 7 1.23609271356527\\
7 8 1.23609271356528\\
7 9 1.23609271356529\\
7 10 1.2360927135653\\
8 3 0.749638072622376\\
8 4 0.934541959616262\\
8 5 1.2360927135653\\
8 6 1.23609271356533\\
8 7 1.23609271356532\\
8 8 1.23609271356531\\
8 9 1.23609271356532\\
8 10 1.2360927135653\\
9 3 0.749638072622402\\
9 4 0.934541959616254\\
9 5 1.23609271356533\\
9 6 1.23609271356535\\
9 7 1.23609271356534\\
9 8 1.23609271356534\\
9 9 1.23609271356535\\
9 10 1.23609271356532\\
10 3 0.749638072622391\\
10 4 0.934541959616241\\
10 5 1.23609271356533\\
10 6 1.23609271356531\\
10 7 1.23609271356532\\
10 8 1.23609271356535\\
10 9 1.23609271356531\\
10 10 1.23609271356534\\
};
\end{axis}
\end{tikzpicture}%
	\caption{Sample variance of $G_3(e^{j\omega}, \hat \theta_3)$ as a function of the number of estimated parameters of $G_1$ and $G_2$.}
	\label{fig:MA}
\end{figure}

Corollary~\ref{cor:variance_increase} is illustrated in Figure~\ref{fig:MA}, where the following systems are identified using $N = 500$ input-output measurements: 
\begin{IEEEeqnarray*}{rClrCl}
\label{eq:example_system}
G_i &=& \tilde \Gamma_i \theta_i, \quad i = 1,2,3
&\qquad 
\tilde \Gamma_i(q) &=& F(q)^{-1} \Gamma_i(q), \\
\Gamma_i(q) &=&
\begin{bmatrix}
 \Bc_1(q), \dotsc, \Bc_{n_i}(q) 
 \end{bmatrix}, &\qquad
 \Bc_k(q) &=& q^{-k},
 \yesnumber
\end{IEEEeqnarray*}
with
\begin{IEEEeqnarray*}{rClrCl}
F(q) &=& \frac{1}{1 - 0.8 q^{-1} } , & \qquad
\theta_1^0 &=& \begin{bmatrix}
1 & 0.5 & 0.7
\end{bmatrix}^T, \\
\theta_2^0 &=& \begin{bmatrix}
1 & -1 & 2
\end{bmatrix}^T, & \qquad
\theta_3^0 &=& \begin{bmatrix}
1 & 1 & 2 & 1 & 1
\end{bmatrix}^T \!\!\! .
\end{IEEEeqnarray*} 
The input $u(t)$ is drawn from a Gaussian distribution with variance $\sigma^2 = 1$, filtered by $F(q)$. The measurement noise is normally distributed with covariance matrix $\Lambda = \Lambda_{CH}\Lambda_{CH}^T$, where
\begin{IEEEeqnarray*}{rCl}
\Lambda_{CH} = \begin{bmatrix}
1 	& 	0	&	0	\\
0.6	&	0.8	&	0	\\
0.7 & 0.7 & 0.1
\end{bmatrix},
\end{IEEEeqnarray*} 
thus $\lambda_1 = \lambda_2= \lambda_3 = 1$.
The sample variance is computed using 
\begin{IEEEeqnarray*}{rCl}
 \cov \hat \theta_{\mathrm{s}} &=& \frac{1}{MC}\sum_{k=1}^{MC} |G_3(e^{j\omega_0},\theta_3^0)- G_3(e^{j\omega_0},\hat \theta_3)|^2,
\end{IEEEeqnarray*} 
where $MC = 2000$ is the number of realizations of the input and noise. 
The same realizations of the input and noise are used for all model orders.

The variance of $G_3(e^{j\omega}, \hat \theta_3)$ increases with increasing $n_i$, $i = 1,2$, but only up to the point where $n_i = n_3 = 5$. After that, any increase in $n_1$ or $n_2$ does not increase the variance of $G_3(e^{j\omega}, \hat \theta_3)$, as can be seen in Figure~\ref{fig:MA}. The behavior can be explained by Corollary \ref{cor:variance_increase}: when $n_3 \ge n_1,n_2 $, $G_3$ is the last block, having the highest number of parameters, and any increase in $n_1,n_2$ increases the variance of $G_3$. When for example $n_1\ge n_3$, the blocks should be reordered so that $G_3$ comes before $G_1$. In this case, when $n_1$ increases the first condition of Corollary \ref{cor:variance_increase} does not hold and hence the variance of $G_3(e^{j\omega}, \hat \theta_3)$ does not increase further.

\subsection{Effect of removing one parameter}
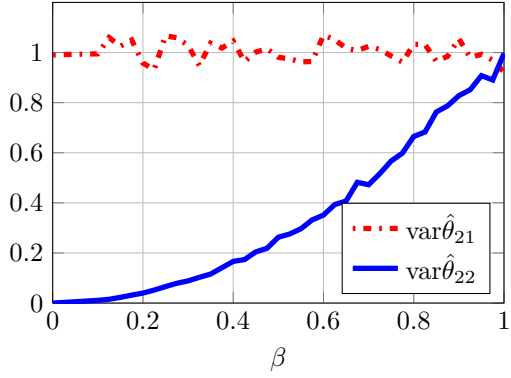
\begin{figure}[!ht]
    \centering    
%
%
\begin{tikzpicture}

\begin{axis}[%
width=\figurewidth,
height=\figureheight,
scale only axis,
xmin=0,
xmax=1,
xtick={  0, 0.2, 0.4, 0.6, 0.8,   1},
xlabel={$\beta$},
xmajorgrids,
ymin=0,
ymax=1.2,
ytick={  0, 0.2, 0.4, 0.6, 0.8,   1},
ymajorgrids,
name=plot1,
legend style={at={(0.97,0.03)},anchor=south east,draw=black,fill=white,legend cell align=left}
]
\addplot [
color=red,
dash pattern=on 1pt off 3pt on 3pt off 3pt,
line width=2.0pt
]
table[row sep=crcr]{
0.001 0.990818435297882\\
0.1 0.994885871596075\\
0.125 1.05978752689978\\
0.15 1.02511546826788\\
0.175 1.05005539444582\\
0.2 0.957649593130851\\
0.225 0.932243680715961\\
0.25 1.06713359101004\\
0.275 1.05902984009893\\
0.3 1.02853009005755\\
0.325 0.948267165454544\\
0.35 1.03734656217939\\
0.375 1.01784790062555\\
0.4 1.0458872727799\\
0.425 0.964155730985572\\
0.45 1.00163121728744\\
0.475 1.01522683057648\\
0.5 0.981236012383618\\
0.525 0.974762215787994\\
0.55 0.962789224991802\\
0.575 0.964178123365545\\
0.6 1.06916979944943\\
0.625 1.05171344913087\\
0.65 1.01724345344588\\
0.675 1.00753045799813\\
0.7 1.02466153822605\\
0.725 1.01240467326144\\
0.75 0.986053089107481\\
0.775 0.961372011845188\\
0.8 1.0332149475958\\
0.825 1.03180687061121\\
0.85 0.965667066333517\\
0.875 0.984646425706888\\
0.9 1.05549805716467\\
0.925 0.980140843817182\\
0.95 0.992907228837606\\
0.975 0.969783076553231\\
1 0.921326973042311\\
};
\addlegendentry{$\mathrm{var} \hat \theta_{21}$};

\addplot [
color=blue,
solid,
line width=2.0pt
]
table[row sep=crcr]{
0.001 1.03520305509892e-06\\
0.1 0.0107267625106279\\
0.125 0.0150504238587857\\
0.15 0.0228157693360597\\
0.175 0.031660139330561\\
0.2 0.0399025074476164\\
0.225 0.0519347869918284\\
0.25 0.0658234937796144\\
0.275 0.0790218074947223\\
0.3 0.0885545871427352\\
0.325 0.102758092578143\\
0.35 0.115792835849245\\
0.375 0.14105236465198\\
0.4 0.166406547789313\\
0.425 0.173527861930704\\
0.45 0.204264144079571\\
0.475 0.218600337663263\\
0.5 0.262914192170222\\
0.525 0.275582920619583\\
0.55 0.296559503892826\\
0.575 0.332096307594496\\
0.6 0.351118540290889\\
0.625 0.393291357891636\\
0.65 0.407165414569293\\
0.675 0.48199109449862\\
0.7 0.472330182217597\\
0.725 0.517023010638669\\
0.75 0.567039086601611\\
0.775 0.598550873513195\\
0.8 0.665091132341268\\
0.825 0.682828734130626\\
0.85 0.762712071953172\\
0.875 0.787322334735176\\
0.9 0.827894476615921\\
0.925 0.851412193261338\\
0.95 0.907760603167109\\
0.975 0.890437408629264\\
1 0.996392893640718\\
};
\addlegendentry{$\mathrm{var} \hat \theta_{22}$};

\end{axis}
\end{tikzpicture}%
	\caption{Sample variance of the parameters of the first module (as functions of $\beta$) when the second module has one parameter.}
	\label{fig:first_experiment}
\end{figure}

In the second set of numerical experiments, we simulate the same type of system as in Section~\ref{subsec:example}. 
We let $\beta$ vary in the interval $[0.001,\,1]$. For each $\beta$, we generate $MC = 2000$ Monte Carlo experiments, where in each of them we collect $N=500$ input-output samples. At the $i$-th Monte Carlo run, we generate new trajectories for the input and the noise and we compute $\hat \theta_i$ as in \eqref{eq:WLS_estimate}. The sample covariance matrix, for each $\beta$, is computed as
\begin{IEEEeqnarray*}{rCl}
\cov \hat \theta_{\mathrm{s}} &=& \frac{1}{MC}\sum_{i=1}^{MC} (\hat \theta_i - \theta)(\hat \theta_i - \theta)^T \! .
\end{IEEEeqnarray*}
Figure \ref{fig:first_experiment} shows that the variance of $\hat \theta_{21}$ is always close to one, no matter what the value of $\beta$ is. It also shows that the variance of the estimate $\hat \theta_{22}$ behaves as $\beta^2$. In particular, when $\beta$ approaches 0 (i.e., almost perfectly correlated noise processes), the variance of such estimate tends to 0.
All of the observations are as predicted by Corollary~\ref{thm:cov_diag} and the example in Section~\ref{subsec:example}.

\subsection{Optimal correlation structure}
A third simulation experiment is performed in order to illustrate Theorem \ref{th:optimal_correlation}. We consider a system with $m=3$ outputs; the modules are
\begin{IEEEeqnarray*}{rClrCl}
G_1(q) & = & 0.3 q^{-1} \! ,
& \quad G_2(q) & = & 0.8 q^{-1} -0.4 q^{-2} \! , \\
G_3(q) & = & 0.1 q^{-1} +0.2 q^{-2} \! ,
\end{IEEEeqnarray*}
so that
$$
\theta = [\theta_{11}\,\theta_{21}\,\theta_{31}\,\theta_{22}\,\theta_{32}]^T = [0.3\,\,\,0.8\,\,\,0.1\,\,\,-0.4\,\,\,0.2]^T\,.
$$
The noise process is generated by the following upper triangular Cholesky factor:
\begin{align*}
R & = \begin{bmatrix} \varepsilon & \sqrt{1-\varepsilon^2}\cos{\alpha} & \sqrt{1-\varepsilon^2}\sin{\alpha} \\
                        0 & 0.8 &  0.6 \\
                        0 &  0  &   1 \\
                       \end{bmatrix} \nonumber  = \begin{bmatrix} \varepsilon & p^T \\ 0  & M  \end{bmatrix} \,,
\end{align*}
where $\varepsilon = 0.1$ and $\alpha \in [0,\,\pi]$ is a parameter  tuning the correlation of $e_1(t)$ with $e_2(t)$ and $e_3(t)$. The purpose of this experiment is to show that, when $\alpha$ is such that $p = [\sqrt{1-\varepsilon^2}\cos{\alpha} \quad \sqrt{1-\varepsilon^2}\sin{\alpha}]^T$ is aligned with the first right-singular vector $v_1$ of $M$, then the total variance of the estimate of the sub-vector $\tilde \theta = [\theta_{12}\,\theta_{22}]^T$ is minimized. In the case under analysis, $v_1 = [0.447\,\,\,0.894]^T = [\cos{\alpha_0}\,\,\sin{\alpha_0}]^T$, with $\alpha_0 = 1.11$. We let  $\alpha$ take values in $[0,\,\pi]$ and for each $\alpha$ we generate $MC = 2000$ Monte Carlo runs of $N=500$ input-output samples each. We compute the sample total variance of $\tilde \theta$ as
\begin{IEEEeqnarray*}{rCl}
\Tvar \;\hat{\tilde \theta}_{\mbox{s}}  
&=& \frac{1}{MC}\sum_{i=1}^{MC} \left((\hat \theta_{22,i} - \theta_{22})^2 + (\hat \theta_{32,i} - \theta_{32})^2 \right) \!,
\end{IEEEeqnarray*}
where $\hat \theta_{22,i}$ and $\hat \theta_{32,i}$ are the estimates obtained at the $i$-th Monte Carlo run.

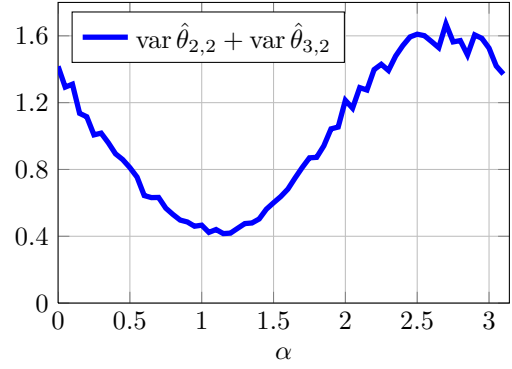
\begin{figure}[!ht]
    \centering    
%
%
\begin{tikzpicture}

\begin{axis}[%
width=\figurewidth,
height=\figureheight,
scale only axis,
xmin=0,
xmax=3.14159265358979,
xtick={0,0.5,1,1.5,2,2.5,3},
xticklabels={0,0.5,1,1.5,2,2.5,3},
xlabel={$\alpha$},
xmajorgrids,
ymin=0,
ymax=1.8,
ytick={  0, 0.4, 0.8, 1.2, 1.6},
ymajorgrids,
legend style={at={(0.03,0.97)},anchor=north west,draw=black,fill=white,legend cell align=left}
]
\addplot [
color=blue,
solid,
line width=2.0pt
]
table[row sep=crcr]{
0 1.41754080415461\\
0.05 1.29307031863212\\
0.1 1.31034418646953\\
0.15 1.1359640103263\\
0.2 1.11550584802069\\
0.25 1.00704798059736\\
0.3 1.01743150367511\\
0.35 0.959672937692806\\
0.4 0.893616052444127\\
0.45 0.859178250146741\\
0.5 0.81100864288756\\
0.55 0.754265540378456\\
0.6 0.643676864209814\\
0.65 0.631474737495688\\
0.7 0.632426860114713\\
0.75 0.568212717877039\\
0.8 0.530358005356942\\
0.85 0.496907806762119\\
0.9 0.485667639985688\\
0.95 0.460126961836715\\
1 0.466299214802827\\
1.05 0.423231608497019\\
1.1 0.440335597450245\\
1.15 0.416032822764961\\
1.2 0.418734285232464\\
1.25 0.447554598444223\\
1.3 0.475884388705693\\
1.35 0.479613687806601\\
1.4 0.503575797336119\\
1.45 0.562029541274954\\
1.5 0.600619377884408\\
1.55 0.637510436061427\\
1.6 0.682637685914398\\
1.65 0.749125399615997\\
1.7 0.812085208390585\\
1.75 0.869625549931935\\
1.8 0.872477047280432\\
1.85 0.941305377927098\\
1.9 1.04284504714328\\
1.95 1.05399078047091\\
2 1.21390924876346\\
2.05 1.16689786969665\\
2.1 1.28987863729493\\
2.15 1.27556580977618\\
2.2 1.39789690414935\\
2.25 1.42958040042146\\
2.3 1.39253659823598\\
2.35 1.47925466545538\\
2.4 1.54186333513845\\
2.45 1.59347548309878\\
2.5 1.60979738682042\\
2.55 1.60097624356092\\
2.6 1.565803528271\\
2.65 1.52927593175675\\
2.7 1.67125136954606\\
2.75 1.56368798308007\\
2.8 1.57148471660814\\
2.85 1.48599166357315\\
2.9 1.60574915860819\\
2.95 1.58502029326213\\
3 1.52632794099588\\
3.05 1.41962081370149\\
3.1 1.37092965865776\\
};
\addlegendentry{$\mathrm{var}\, \hat \theta_{2,2} + \mathrm{var}\, \hat \theta_{3,2}$};

\end{axis}
\end{tikzpicture}%
	\caption{Total sample variance of the parameter vector $\tilde \theta$ as function of $\alpha$.}
	 \label{fig:second_experiment}
\end{figure}

The results of the experiment are reported in Figure \ref{fig:second_experiment}. It can be seen that the minimum total variance of the estimate of $\hat \theta$ is attained for values close to $\alpha_0$ (approximations are due to low resolution of the grid of values of $\alpha$). An interesting observation regards the value of $\alpha$ for which the total variance is maximized: this happens when $\alpha = 2.68$, which yields  the second right-singular vector of $M$, namely $v_2 = [-0.894\,\,\,0.447]^T$.

\section{Conclusions}
\label{sec:conclusions}

The purpose of this paper has been to examine how the estimation accuracy of a linear SIMO model depends on the correlation structure of the noise and model structure and model order. A formula for the asymptotic covariance of the frequency response function estimate and the model parameters has been developed for the case of temporally white, but possibly spatially correlated additive noise. 
It has been shown that the variance decreases when parts of the noise can be linearly estimated from measurements of other blocks with less estimated parameters.  
The expressions reveal how the order of the different blocks and the correlation of the noise affect the variance of one block. In particular, it has been shown that the variance of the block of interest levels off when the number of estimated parameters in another block reaches the number of estimated parameters of the block of interest. 
The optimal correlation structure for the noise was determined for the case when one block has one parameter less than the other blocks. 

\bibliographystyle{automatica}  
\bibliography{../../Common_bib_files/network_systems_2} 


\appendix

\section{Proof of Lemma~\ref{lem:variance_non_estimable_part_of_noise} }
\label{sec:proof_variance_non_estimable_part_of_noise}
Let $v(t)= \Lambda_{CH}^{-1}e(t)$ for some real valued random variable $e(t)$ ($\Lambda_{CH}^{-1}$ exists and is unique for $\Lambda > 0$ \citep{horn90}). Then $\cov v(t) = I$. 
Similarly $e(t) = \Lambda_{CH} v(t)$. 
The set $\{ v_1(t), \dotsc, v_{j}(t)\}$ is a function of $e_1(t), \dotsc, e_{j}(t)$ only and vice versa, for all $1 \le j\le m$. Thus,
if $e_1(t), \dotsc, e_{j}(t)$ are known, then also $\{ v_1(t), \dotsc, v_{j}(t)\}$ are known, but nothing is known about $\{ v_{j+1}(t), \dotsc, v_{m}(t)\}$. Thus, for $j < i$ 
the best linear estimator of $e_i(t)$ given $e_1(t), \dotsc, e_{j}(t)$, is 
\begin{IEEEeqnarray*}{rCl}
\label{eq:noise_estimate}
\hat e_{i|j}(t) = \sum_{k=1}^j \gamma_{ik} v_k(t),
\yesnumber
\end{IEEEeqnarray*}
and 
\begin{IEEEeqnarray*}{rCl}
e_i(t)-\hat e_{i|j}(t)
\end{IEEEeqnarray*}
has variance
\begin{IEEEeqnarray*}{rCl}
\lambda_{i|j} = \sum_{k=j+1}^i \gamma_{ik}^2  .
\end{IEEEeqnarray*} 
For the last part of the lemma, we realize that the dependence of $\hat e_{i|j}(t)$ on $e_j(t)$ in Equation~\eqref{eq:noise_estimate} is given by $\gamma_{ij}/\gamma_{jj}$ (since $v_1(t),\dotsc,v_{j-1}(t)$ do not depend on $e_j(t)$). Hence $\hat e_{i|j}(t)$ depends on $e_j(t)$ if and only if $\gamma_{ij} \neq 0$.

\section{Proof of Theorem~\ref{cor:variance_G_i} }
\label{sec:proof_variance_G_i}

Before giving the proof of Theorem~\ref{cor:variance_G_i} we need the following auxiliary lemma.

\begin{lem}
\label{lem:proj_chol}
Let $\Lambda > 0$ and real and its Cholesky factor $\Lambda_{CH}$ be partitioned according to $e_{1:\chi_k-1}$ and $e_{\chi_k:m}$,  
\begin{IEEEeqnarray*}{rClrCl}
\Lambda &=& 
\begin{bmatrix}
\Lambda_{1} & \Lambda_{12} \\
\Lambda_{21} & \Lambda_{2}
\end{bmatrix},& 
\qquad \Lambda_{CH} &=&
\begin{bmatrix}
(\Lambda_{CH})_{1} & 0 \\
(\Lambda_{CH})_{21} & (\Lambda_{CH})_{2}
\end{bmatrix}. \IEEEeqnarraynumspace
\end{IEEEeqnarray*}
Then 
\begin{IEEEeqnarray*}{rCl}
\Lambda_{\chi_k:m|\chi_k-1} &=& (\Lambda_{CH})_{2}(\Lambda_{CH})_{2}^T.
\end{IEEEeqnarray*}
\end{lem}
\begin{pf}
By the derivations of Lemma~\ref{sec:proof_variance_non_estimable_part_of_noise}, for some $v(t)$ with $\cov v(t) = I$, $e(t) = \Lambda_{CH} v(t)$ and $v_{1:\chi-1}(t)$ are known since $e_{1:\chi-1}(t)$ are known. Furthermore $\hat{e}_{\chi_k:m|\chi_k-1}(t) = (\Lambda_{CH})_{21} v_{1:\chi-1}(t)$, which implies $e_{\chi_k:m|\chi_k-1}(t) - \hat{e}_{\chi_k:m|\chi_k-1}(t) =  (\Lambda_{CH})_{2}v_{\chi_k:m}(t)$ and the results follows since $\cov v_{\chi_k:m}(t) = I$. $\hfill \blacksquare$
\end{pf}

The asymptotic variance is given by \eqref{eq:inp_psi} with
\begin{IEEEeqnarray*}{rCl}
 \Psi(q) &=& \tilde \Psi(q) \Lambda_{CH}^{-T}.
\end{IEEEeqnarray*} 
Let $n = n_1 + \dotsb + n_m$. 
From the upper triangular structure of $\Lambda_{CH}^{ -T}$ and $n_1 \le n_2 \le \dotsc \le n_m$, an orthonormal basis for $\Sc_{\Psi}$, the subspace spanned by the rows of $\Psi$, is given by
\begin{IEEEeqnarray*}{rClrcl}
\!\! \Bc_k^\Sc(e^{j\omega}) & \defeq & 
\begin{bmatrix}
\Bc_k & 0 & \dotsc & 0
\end{bmatrix}\! , & 
k &=& 1, \dotsc ,n_1
\\
\!\! \Bc_k^\Sc(e^{j\omega}) & \defeq & 
\begin{bmatrix}
0\! & \Bc_{k-n_1} \! & 0\! & \dotsc \!
\end{bmatrix}\! , & \:\:
 k &=& n_1+1, \dotsc ,n_2  
 \\
 & \vdots & \IEEEyesnumber
 \\
\!\! \Bc_k^\Sc(e^{j\omega}) & \defeq &
\begin{bmatrix}
 \dotsc \! & 0 \! &\Bc_{k-n+n_m}  \!
\end{bmatrix} \! , &
 k &=& n-n_m \! + \! 1, \dotsc ,n.   
\end{IEEEeqnarray*}  
First note that
\begin{IEEEeqnarray*}{rCl}
 \frac{\partial G}{\partial \theta} = \Psi \Lambda_{CH}^T.
\end{IEEEeqnarray*} 
Then, using Theorem~\ref{lem:ascov_as_sum_of_basis_functions},
\begin{IEEEeqnarray*}{rCl}
\sigma^2 \, \asvar \hat G
&=& \Lambda_{CH} \sum_{k=1}^n \Bc^\Sc_k(e^{j\omega_0})^* \Bc^\Sc_k(e^{j\omega_0})  \Lambda_{CH}^T .
\IEEEeqnarraynumspace
\end{IEEEeqnarray*} 
Sorting the sum with respect to the basis functions $\Bc_k(e^{j\omega_0})$, we get
\begin{IEEEeqnarray*}{rCl}
\sigma^2 \,\ascov{\hat G_i } &=&  \Lambda_{CH} \sum_{k=1}^{n_m}  |\Bc_k(e^{j\omega_0})|^2
\begin{bmatrix}
0_{\chi_k-1} & 0 \\
0 & I
\end{bmatrix}
\Lambda_{CH}^T . \IEEEeqnarraynumspace
\end{IEEEeqnarray*} 
Using Lemma~\ref{lem:proj_chol}
\begin{IEEEeqnarray*}{rCl}
\ascov{\hat G } &=& \frac{1}{\sigma^2} \sum_{k=1}^{n_m}  
\begin{bmatrix}
0_{\chi_k-1} & 0 \\
0 & \Lambda_{\chi_k:m|\chi_k-1}
\end{bmatrix}
|\Bc_k(e^{j\omega_0})|^2. \IEEEeqnarraynumspace
\end{IEEEeqnarray*} 
Thus \eqref{eq:thm_ascov_G} follows. We now show the first part of the theorem, that
\begin{IEEEeqnarray*}{rCl} 
 	\ascov \hat{ \bar{\theta}}
 	&=&  \frac{1}{\sigma^2}\diag (\Lambda_{1:m}, \Lambda_{\chi_2:m|\chi_2-1}, \dotsc,  \Lambda_{\chi_{n_m}:m|\chi_{n_m}-1 } ).
\end{IEEEeqnarray*}
The covariance of $\hat G$ can be expressed as
\begin{IEEEeqnarray*}{rCl}
\label{eq:cor_ascov_bar_theta}
\ascov \hat G &=& T \ascov \hat{ \bar{ \theta}} \:   T^*
\yesnumber
\end{IEEEeqnarray*}
where 
\begin{IEEEeqnarray*}{rCl}
T &=& \begin{bmatrix}
\Bc_1 I(1) & \Bc_2 I(2) \dotsc & \Bc_{n_m} I(n_m)
\end{bmatrix},
\\
I(k) &=& \begin{bmatrix} {\bf{0}} \\  	I_{m-\chi_k+1} \end{bmatrix}\in \Rb^{m \times (m-\chi_k+1)}.
\end{IEEEeqnarray*}
However, \eqref{eq:cor_ascov_bar_theta} equals \eqref{eq:thm_ascov_G} for all $\omega$ and the theorem follows. 

%

\section{Proof of Corollary~\ref{cor:variance_increase} }
\label{sec:proof_variance_increase}
To prove Corollary~\ref{cor:variance_increase} we will use \eqref{eq:ascov_G_i_hat}. First, we make the assumption that $j$ is the last module that has $n_j$ parameters. This assumption is made for convenience since reordering all modules with $n_j$ estimated parameters does not change \eqref{eq:ascov_G_i_hat}. First of all, we see that if 
\begin{IEEEeqnarray*}{rCl}
n_j \ge n_i,
\end{IEEEeqnarray*} 
then \eqref{eq:ascov_G_i_hat} does not increase when $n_j$ increases. 
If instead 
\begin{IEEEeqnarray*}{rCl}
n_j < n_i,
\end{IEEEeqnarray*} 
the increase in variance is given by 
\begin{IEEEeqnarray*}{rCl}
\gamma_{ij}^2|\Bc_{n_j+1}(e^{j\omega_0})|^2,
\end{IEEEeqnarray*} 
which is non-zero iff $\gamma_{ij} \neq 0$ and $|\Bc_{n_j+1}(e^{j\omega_0})|^2 \neq 0$. 
From Lemma~\ref{lem:variance_non_estimable_part_of_noise} the theorem follows.

%

\section{Proof of Lemma \ref{lem:proof_lemma_inversion}} \label{sec:proof_lemma_inversion}

The inverse of $B$ is
\begin{IEEEeqnarray*}{rCl}
B^{-1} &=& \begin{bmatrix} \eta^{-1} & - q^T \\ 0 & M^{-1} \end{bmatrix}, \quad\, q^T \defeq \eta^{-1} p^T M^{-1} \!,
\end{IEEEeqnarray*}
so that 
\begin{IEEEeqnarray*}{rCcCl}
\Lambda^{-1} &=& B^{-T}B^{-1} &=& \begin{bmatrix} \eta^{-2} & - q^T \eta^{-1} \\ - q\eta^{-1} &  M^{-T}M^{-1} + q q^T   \end{bmatrix} .
\end{IEEEeqnarray*}
Hence, using the Sherman--Morrison formula \citep{Bartlett1951}
\begin{IEEEeqnarray*}{rCl}
\Lambda_{2:m|1} & = & \left( M^{-T}M^{-1} + q q^T \right)^{-1} \\
                & = & MM^T - \frac{1}{k}  MM^T qq^T MM^T \\
                & = & MM^T - \frac{1}{k}  \frac{M p p^T M^T}{\eta^2} \,,
\end{IEEEeqnarray*}
where
\begin{IEEEeqnarray*}{rCl}
k & = & 1 + q^T MM^T q = 1 + \frac{p^Tp}{\eta^2} \nonumber \\
  & = & \frac{\eta^2 + p^Tp}{\eta^2} = \frac{\|b_1\|_2^2 }{\eta^2} \nonumber \\
  & = & \frac{\lambda_1}{\eta^2} \,,
\end{IEEEeqnarray*}
so \eqref{eq:one_less_parameter} follows.


\end{document}